# Investigating molecular crowding during cell division in budding yeast with FRET

Sarah Lecinski[1,a], Jack W Shepherd[1,a,b], Lewis Frame[c], Imogen Hayton[b], Chris MacDonald[b], Mark C Leake[a,b]*

[1]These authors contributed equally

[a] Department of Physics, University of York, York, YO10 5DD

[b] Department of Biology, University of York, York, YO10 5DD

[c] School of Natural Sciences, University of York, York, YO10 5DD

* To whom correspondence should be addressed. Email mark.leake@york.ac.uk

**Abstract**

Cell division, aging, and stress recovery triggers spatial reorganization of cellular components in the cytoplasm, including membrane bound organelles, with molecular changes in their compositions and structures. However, it is not clear how these events are coordinated and how they integrate with regulation of molecular crowding. We use the budding yeast *Saccharomyces cerevisiae* as a model system to study these questions using recent progress in optical fluorescence microscopy and crowding sensing probe technology. We used a Förster Resonance Energy Transfer (FRET) based sensor, illuminated by confocal microscopy for high throughput analyses and Slimfield microscopy for single-molecule resolution, to quantify molecular crowding. We determine crowding in response to cellular growth of both mother and daughter cells, in addition to osmotic stress, and reveal hot spots of crowding across the bud neck in the burgeoning daughter cell. This crowding might be rationalized by the packing of inherited material, like the vacuole, from mother cells. We discuss recent advances in understanding the role of crowding in cellular regulation and key current challenges and conclude by presenting our recent advances in optimizing FRET-based measurements of crowding whilst simultaneously imaging a third color, which can be used as a marker that labels organelle membranes. Our approaches can be combined with synchronised cell populations to increase experimental throughput and correlate molecular crowding information with different stages in the cell cycle.

**Abbreviations:** FRET: Förster Resonance Energy Transfer; PM: plasma membrane; ConA: Concanavalin A; OD: optical density; YPD: Yeast Extract-Peptone-Dextrose medium; SD: synthetic defined medium; DIC: differential interference contrast

Introduction

The term "molecular crowding" describes the range of molecular confinement-induced effects (e.g. mobility, soft attraction, and repulsion forces) observed in a closed system of concentrated molecules. Cells are highly crowded membrane-bound environments containing a range of biomolecular species including proteins, polysaccharides and nucleic acids. Typically, these molecules occupy a huge volume of the cell (up to 40%), equivalent to a concentration of up to 400 mg/mL (Fulton, 1982; Zimmerman et al., 1991). Two terms can be encountered in the literature: "macromolecular crowding" referring to dynamic effects of volume exclusion encountered between two molecules and "macromolecular confinement" referring to the same effect caused by the static shape and size of the system (Huan-Xiang et al., 2008; Sanfelice et al., 2013; Zhang et al., 2019). Both describe a typical free-space limitation occurring in a highly concentrated environment of molecules which leads to non-specific interactions between macromolecules in close proximity (Sarkar et al., 2013). Excluded volume theory is a key concept to understanding macromolecular crowding (Garner & Burg, 1994; Kuznetsova et al., 2014). By their presence, molecules exclude access to the solvent/surface of other molecules. This imposed volume restriction where exclusion is dependent on the molecule's size and shape (Fig. 1). As a result, if each molecule excludes a certain volume from every other, and the mobility of each is also reduced – a molecule can only diffuse into an available volume, which effectively slows the time scale of the overall diffusive process. Note also that the overall effect of the multiple biomolecular species *in vivo* is referred to exclusively as crowding with the term "concentration" used generally for individual protein species, which may be at a low concentration in an overall high crowding environment or *vice versa* (Ellis, 2013; Minton, 2006).

An extensive number of studies on molecular crowding have been conducted *in vitro* mimicking the cytoplasmic crowded environment (Benny & Raghunath, 2017; Davis et al., 2020) using inert and highly water-soluble crowding agents such as glycerol (Y. L. Zhou et al., 2006), PEG (Benny & Raghunath, 2017), and Ficoll (Christiansen et al., 2013; B. Van Den Berg et al., 1999). Several of these studies have described the role of crowding in protein stability (König et al., 2021) and its impact on the free-energy of folding (Minton, 2006; Nguemaha et al., 2019) and kinetics of interaction (Phillip & Schreiber, 2013; Stagg et al., 2007). *In vivo*, dynamics of repulsion and attraction forces between macromolecules has been described as one of the effects of crowding (Garner & Burg, 1994), with a role on molecular rearrangement (Rivas et al., 2004), promoting oligomerization and aggregation (Christiansen et al., 2013). Interestingly, in theory the excluded volume of two monomers does not overlap, but if molecules can biochemically interact to form a stable new entity, this new molecule will now have its own excluded volume changing the solvent availability for other proteins (Allen P, 2001; André & Spruijt, 2020; Poland, 1998) - see Fig. 1). These dynamic effects have raised new interest in the context of protein aggregation (Jing et al., 2020), intracellular organization (Löwe et al., 2020), and membrane-less compartments in the cytoplasm, for example the formation of liquid-liquid phase separation in the cytoplasm (Franzmann et al., 2018; Jin et al., 2021; Park et al., 2020). More generally, macromolecular crowding has been linked to a number of biological process such as protein-protein interaction (Bhattacharya et al., 2013a), protein conformational changes (Dong et al., 2010), stability and folding (Despa et al., 2006; Stagg et al., 2007; Tokuriki, 2004), promoting signalling cascades (Rohwer et al., 1998), regulating biomolecular diffusion (Tabaka et al., 2014; Tan et al., 2013), and intracellular transport (Nettesheim et al., 2020).

Increased crowding in the cytoplasm has also been associated with promoting binding interaction (Bhattacharya et al., 2013b) and complexation with enzymatic reactions (Akabayov et al., 2013), and enhancing gene expression (Hancock, 2014; Mager & Siderius, 2002; Shim et al., 2020) and protein phosphorylation (Ouldridge & Rein ten Wolde, 2014). However, a strong shift in cellular crowding may also have the opposite effect to lower cellular metabolic activities with gene expression and entry to cell division compromised (André & Spruijt, 2020; Ma & Nussinov, 2013). Such metabolic reactions are typically encountered in cells exposed to acute stress. External stress perturbs the enclosed cell and triggers physical, morphological, and metabolic changes (Campisi & D'Adda Di Fagagna, 2007; Canetta et al., 2006; Marini et al., 2018). In general, cellular stress responses are mediated by complex signalling pathways that involve a coordinated set of biomolecular reactions and interactions that modify gene expression, metabolism, protein localization, and in some cases cell morphology. Taken together these determine the cell's ultimate fate, either recovery over time or programmed cell death (Fulda et al., 2010; Galluzzi et al., 2018) (see Fig. 2A).

In general, cellular stresses may be sorted into two categories, either environmental such as for heat shock or nutrient starvation; or of an intracellular nature due to for example toxic by-products of metabolism produced due to aging (Lumpkin et al., 1985; Sinclair & Guarente, 1997), the presence of reactive oxygen species (ROS) in the cytosol (Liochev, 2013; Santos et al., 2018), or the formation of protein aggregates (Groh et al., 2017; López-Otín et al., 2013). Stresses of these types are encountered by the majority of cells populations, which could explain the highly conserved nature of response mechanisms (Kültz, 2020; Lockshin & Zakeri, 2007). Indeed, response to thermal stress mediated by heat shock proteins and associated chaperones, conserved between prokaryotes and even complex eukaryotic cells, is a perfect example (Z. Li & Srivastava, 2003; Verghese et al., 2012). Other highly conserved mechanisms include DNA damage repair systems (essential to preserve cell integrity), e.g. the case of the highly conserved Rad52 protein which operates via the Mre11-Rad50-Nbs1 (MRN) complex to effect double strand break repair (Syed & Tainer, 2018) (arising for example from UV stress (Cadet et al., 2005)) and homologous recombination which is essential for cell integrity (Hakem, 2008; Lamarche et al., 2010).

Upon osmotic stress, when cells are exposed to a high ionic strength environment, the cell volume suddenly reduces because of the osmotic pressure generated, which triggers a diffusion-driven water exchange from the cell cytosol to the external environment. This drastic volume reduction spatially confines the pool of macromolecules inside the cells and directly increases molecular crowding (see Fig. 2B). In response, cells activate biochemical pathways such as that involving the protein Hog1 to produce a gain in volume via internal pressure generation (Tamás et al., 2000) - this occurs notably via the production of glycerol (Hohmann, 2002) and the regulation of transporters at the plasma membrane such as aquaporins to control the glycerol/water ratio (Stefan Hohmann, 2015; Saito & Posas, 2012). In eukaryotic cells, osmostress and osmoregulation have been extensively studied using yeast as a model (Babazadeh et al., 2017; Gibson et al., 2007; Hsu et al., 2015; Özcan & Johnston, 1999). As a model eukaryote (Tissenbaum & Guarente, 2002), yeast has many advantages: it is easy to manipulate, rapid to culture (Petranovic et al., 2010), and as a unicellular organism is particularly susceptible to osmotic shock – the entire yeast cell is exposed to the media as opposed to only a fraction of cells in a multicellular model. Moreover, yeast genetics and essential metabolic pathways are conserved within the eukaryotic kingdom (Duina et al., 2014; Dujon, 1996) and have been extensively studied (Foury, 1997; He et al., 2018).

In *Saccharomyces cerevisiae* (budding yeast) cells, mitosis and cytokinesis take place via a budding process, where a bud emerges and grows from the mother cell (Hsin Chen et al., 2011; Juanes & Piatti, 2016). This process requires the establishment of a cell polarity between the mother cell and daughter cell (Bi & Park, 2012) and therefore close regulation of the cytoplasmic content and spatial organization (Nasmyth, 1996), including changes in membrane morphology. Cell division is therefore a highly asymmetrical process (Higuchi-Sanabria et al., 2014) associated with a range of biochemical processes involving protein-protein interaction (Hsin Chen et al., 2011) and significant material and compartment transport (Champion et al., 2017; Yeong, 2005). All these processes rely on biomolecular rearrangement and volume changes, and as such may impact local macromolecular crowding conditions. Cell growth is often inhibited during stress (X. Li & Cai, 1999; Taïeb et al., 2021), correlated to diversion of metabolic resources diverted to mitigate the resultant detrimental physiological effects (Bonny et al., 2020; Chasman et al., 2014). Growth capability is therefore a key parameter in measuring cellular recovery, with those that fail to reverse the arrest and re-enter a replicative cycle generally dying (Petranovic & Ganley, 2014). Yeast, however, can enter a semi-dormant "quiescent" state, in which cells don't divide and maintain only minimal metabolic functions (Nurse & Broek, 1993). In essence, this is a survival state to economize resources when exposed to prolonged periods of starvation (Valcourt et al., 2012). The cytoplasm of this new state has been shown to transition from a liquid-like to a solid-like state, with an associated acidification of the environment which also results in lower diffusion rates (Munder et al., 2016).

The budding process in *S. cerevisiae* is specifically characterised by the formation of a septin ring on the cell membrane early in the replication cycle (Byers & Goetsch, 1976; Hsin Chen et al., 2011). This septin ring acts as a junction between the mother cell and the daughter cell (Vrabioiu & Mitchison, 2006) and defines the mother cell/daughter cell polarity (Juanes &

Piatti, 2016) (Fig. 3). During this process, the role of the plasma-associated GTPase protein Cdc42 is key to initiate the polarization process (Okada et al., 2013) and trigger recruitment of proteins to form the septin ring at the interface between mother and daughter cell in a region generally known as the bud neck (Faty et al., 2002; McMurray et al., 2011). Several septins are recruited to make this hetero-oligomeric structure, with key proteins Shs1, Cdc3, Cdc10, Cdc11, Cdc12 forming a characteristic double ring shape (McMurray & Thorner, 2009; Vrabioiu & Mitchison, 2006). Meanwhile, the cytoskeleton network is restructured such that it is polarized along the axis of the bud neck between mother and daughter cell (Moseley & Goode, 2006) to facilitate active transport of complexes and organelles (Juanes & Piatti, 2016; Warren & Wickner, 1996) to the daughter cell, mainly accomplished by myosin transport along actin cables (Knoblach & Rachubinski, 2015; Macara & Mili, 2008).

During mitosis essential cellular complexes and molecules travels from the mother cell to the daughter cell and are inherited progressively (Cramer, 2008; Yaffe, 1991). This involves the selection of biomolecules and molecular complexes to travel from mother to daughter, especially in the case of key structures the daughter cell cannot synthesize *de novo* including metabolic compartments such as the vacuole, mitochondria, nucleus, and endoplasmic reticulum (Menendez-Benito et al., 2013; Warren & Wickner, 1996). A recent study (Li et al., 2021) reported organelles inherited in a specific and ordered sequence during cell division. Endoplasmic reticulum (ER) and peroxisomes being transported first in the newly formed bud, followed by the vacuole and mitochondria while the bud grow, and at last, in large buds near cytokinesis, the nucleus is transported to the daughter cell. Decoupling cell cycle progression and bud growth, they showed an overall maintained order and timing of migration, supporting a multi-factorial model for organelle inheritance, thus only partially controlled by metabolic pathways associated to cell division. Also, age- or stress-damaged molecules such as extrachromosomal rDNA circles (Lumpkin et al., 1985; Sinclair & Guarente, 1997) and damaged proteins (Chiti & Dobson, 2017) are excluded from trafficking and remain in the mother cell (Higuchi-Sanabria et al., 2014; López-Otín et al., 2013; C. Zhou et al., 2014). The bud-neck region has been described as a "diffusion barrier" influencing trafficking events and the relative cytosolic volume between mother cell and daughter cell (Clay et al., 2014; Gladfelter et al., 2001; Shcheprova et al., 2008; Sugiyama & Tanaka, 2019; Valdez-Taubas & Pelham, 2003) although given the active super-diffusive transport which dominates trafficking, the existence of a meaningful physical diffusion barrier is still an open debate (Nyström & Liu, 2014; C. Zhou et al., 2011).

The influence of macromolecular crowding at the highly localized region between the mother and daughter cell interface remains unknown as well. Quantifying crowding dynamics – or other physicochemical properties – in living cells is challenging, but recent improvements in optical microscope technology and the development of synthetic fluorescent protein sensors have enabled whole cell measurements of crowding during aging and under osmotic stress (Mouton et al., 2019). In essence, these sensors consist of a FRET-pair of fluorescent proteins on an alpha-helical "spring" (Boersma et al., 2015). In crowded conditions, the proteins are pushed closer together, and the FRET efficiency increases, while in less crowded conditions the FRET efficiency is reduced (Fig. 4A). By imaging with dual-color fluorescence microscopy, the FRET efficiency can be quantified and used as a signature for molecular crowding (Fig. 4B). In general, the DNA coding for these sensors is transformed into the cell by homologous recombination (Liu et al., 2017; Mouton et al., 2019), are expressed endogenously, and localize largely to the cytoplasm with both the vacuole in yeast and plasma membrane being excluded volumes (Mouton et al., 2020; Shepherd et al., 2020a). While some sensors, notably the crGE mCerulean3-mCitrine crowding sensor are not ideal for super-resolution localization microscopy (SMLM) due to photophysical limitations, new generations of crowding sensors such as crGE2.3 based on mEGFP and mScarlet-I have fluorescent proteins which undergo FRET and are separately super-resolvable through techniques such as Slimfield microscopy (Lenn & Leake, 2016). Slimfield uses a bespoke microscope similar to a standard epifluorescence setup, but with a small lateral excitation field (Chiu & Leake, 2011; Lenn & Leake, 2012; Plank et al., 2009; Wollman et al., 2016). The laser beam generates a Gaussian intensity profile over the sample with the field of excitation reduced to a small area of effective width <10 µm (full width at half maximum). The excitation intensity is strongly increased as a result compared to conventional epifluorescence excitation, up to ~100 times comparing to widefield image intensities, since the area targeted is smaller by approximatively a factor 10. This feature enables single-molecule detection at very high speeds, helped by precision localization through iterative Gaussian masking of detected bright spots (Shepherd, Higgins, et al., 2021b). Typically, a stack of images is acquired, first the intense illumination excites typically a large number of fluorophores. These fluorophores are then stochastically photobleached in a stepwise fashion, and it is this step size that is characteristic of the brightness of a single fluorophore in that specific microenvironment that ultimately enables us to determine the stoichiometry of fluorescently-labelled molecular complexes that may comprise multiple copies of the same type of fluorophore. To overcome diffusion in the cell cytoplasm the fluorescence signal needs to be captures within a short sampling window, and we typically use a camera exposure times of ~ 5 ms. This position of single molecules emitting fluorescence can be resolved via Gaussian fitting to a precision that is better than the optical resolution limit, typically a few tens of nm. Dual-color imaging can also be performed with this technology, opening possibilities for performing colocalization experiments as well as single-molecule FRET quantification for individual sensors localized to a super-resolved spatial precision to map molecular crowding throughout the cell.

Here, we implemented Slimfield using a narrowfield illumination mode (Wollman & Leake, 2016), based around a standard epifluorescence microscope with high excitation powers at the sample plane in excess of 1800 W/cm$^2$ which we have demonstrated previously may be used for single molecule localization (Miller et al., 2015). This high intensity excitation promotes excitation of individual fluorophores even at very rapid millisecond sampling time scales, and thus enables very rapid single-molecule tracking in live cells relevant to diffusion not only in relatively viscous membranes but also in the cell cytoplasm for which traditional widefield methods are too slow (Fig. 4C). Typically, the field of view is confined to the area encompassing just a few individual cells to allow rapid readout from the camera pixel array, allowing exploration of millisecond time scale cellular dynamics and accurate diffusion coefficient calculation (Plank et al., 2009; Wollman et al., 2016). Alongside this, tracking single molecules with Slimfield microscopy has been used to resolve oligomerization states of molecular assemblies *in vivo* (Badrinarayanan et al., 2012; Laidlaw et al., 2021; Shashkova et al., 2017, 2020; Sun et al., 2019; Syeda et al., 2019; Wollman et al., 2017). The crGE sensor copy numbers (i.e., number of sensor molecules present per cell) were also measured in previous work (Shepherd et al., 2020), demonstrating that crowding readout in the cell was independent of local sensor concentration. Single crGE molecules were also tracked through the low peak emission intensity of mCerulean3 this had precluded single-molecule FRET measurement.

In our present work here, we present novel methods for quantifying subcellular crowding differences between mother and daughter cells during the *S. cerevisiae* budding process, that have particular relevance to the cell dependent changes exhibited in membrane morphologies. We begin by comparing the average molecular crowding between mother and daughter cells using confocal microscopy of the crGE crowding FRET sensor, and correlate these with cell area. We then use deep learning segmentation and a bespoke semi-automatic Python 3 image analysis framework to quantify subcellular regions along and parallel to the cell division axis. We also quantify crowding during cell recovery from osmotic stress and during cell growth, as well as quantifying crowding FRET readout at the bud-neck region around which the plasma membranes of mother and daughter cells are very tightly pinched. We additionally present a new imaging method with combining crowding sensing with mCerulean3-mCitrine FRET sensor and FM4-64, a fluorescent dye that labels endolysosomal lipids (Vida & Emr, 1995), allowing simultaneous visualization of the vacuole membrane. Finally, we present our most recent data in which we have successfully quantified single molecule FRET *in vivo* using the crGE2.3 crowding sensor and our super-resolution image analysis software suite PySTACHIO (Shepherd, Higgins, et al., 2021a).

**Results and discussion**

**Macromolecular crowding stability between mother cell and daughter cells in budding yeast.**

To investigate if polarization and asymmetry between mother cells and budding daughter cells in dividing *S. cerevisiae* is reflected in the crowding environment we expressed the crGE crowding sensor and grew cells to mid-log phase ($OD_{600} = 0.4 - 0.6$) in synthetic complete media containing glucose as previously described (Mouton et al., 2020; Shepherd et al., 2020a) and performed confocal microscopy. Fig. 5 shows the average ratiometric FRET characterized between mother cells and daughter cells/growing buds. This analysis showed a similar FRET readout for mother cells and daughter cells (Fig. 5A and B - Supplemental data. Statistic table 1A). We found that for exponentially dividing cells, the individual cellular area was not correlated with FRET (Fig. 5C and D), consistent with the essential role of crowding stability in cellular integrity and survival (Mouton et al., 2020; J. Van Den Berg et al., 2017). However, there was a greater range of ratiometric FRET values found for smaller cells, which includes the growing daughter cells and buds (Fig. 5C). We hypothesize that this is due to a lower cell volume leading to smaller cells having a greater sensitivity to their immediate environment during initial growth. Moreover, in growing/budding cells the exact stage in the replication cycle may be correlated with crowding, as transport of large material such as inherited organelles may lead to relatively high short-term crowding variability.

Figure 2D shows two sub-populations of budding yeast plotted against bud area. We have split the analysis into two groups, the first displaying higher ratiometric FRET in the mother cell than in the daughter cell and the second where lower ratiometric FRET was measured in the mother cell than in the daughter cell. For those two categories, we see that cell area between the two conditions is statistically equivalent with no statistical difference as measured by both Student's *t*-test and the non-parametric Brunner-Munzel test between the two populations (Supplemental data. Statistic table. 1B). We therefore conclude that the cell size during normal growth is not a predictive factor for subcellular crowding.

**Local crowding readout and cell-by-cell tracking for timeline response to cellular growth and osmotic stress**

To investigate how crowding progresses through the whole replicative cycle we performed confocal imaging with cells immobilized with Concanavalin A (Con A) on a bespoke microfluidics system (Laidlaw et al., 2021) and performed time lapse experiment on growing cells. Analysis was performed using a bespoke Python 3 utility that generated FRET/mCerulean3 ratiometric heat maps showing local regions of high FRET intensity. Figure 6A shows ratiometric heat maps of cells that have undergone 1M NaCl osmotic shock after 20 min in media lacking salt (Fig. 6A - left panel) and ratiometric maps of a cell budding in standard growth conditions (Fig. 6A - right panel). In both conditions, we see a heterogeneous distribution of values across the cytoplasmic volume, and an overall increase of crowding values when 1M NaCl shock is introduced. We also qualitatively observe low and high localized regions of crowding values in budding cells in the absence of osmotic stress (Fig. 6 - white arrows highlighting hot-spot regions).

After image registration and heat map analysis, cell segmentation was performed using the YeastSpotter deep learning model (Lu et al., 2019), and whole cell ratiometric FRET values were individually tracked using a simple centroid tracking method to plot cell crowding through time. Figure 6B shows cell-by-cell tracking analysis, where the mean FRET efficiency for each cell is plotted against time. We see that in the shocked population crowding rises sharply shortly after stress media is introduced, but over time the cells recover and the ratiometric FRET reduces to and beyond its initial value (Fig. 6B - on the left). Meanwhile, ratiometric FRET increases slowly in non-stress media which we hypothesize is associated with the replicative cycle of *S. cerevisiae* (Fig. 6B - on the right).

**Macromolecular crowding in the region of the bud neck**

To further investigate local crowding dynamics during cell division, we focussed on the bud neck, as the narrow region connecting mother cell and daughter cell during mitosis, key to establishment polarity and molecular traffic between the two cells (Faty et al., 2002; Perez &

Thorner, 2019). We selected several markers of the bud neck (Myo1, Cdc1, Cdc12 and Hof1) tagged with super-folder GFP (sfGFP) and determined their localization to this region (Weill et al., 2018) by confocal microscopy, and confirmed that the tag did not perturb growth (Supplemental Fig. S1A ,S1B). We observed ring formation along bud emergence and the splitting event at the end of division for cells expressing sfGFP-Hof1 (Fig. 7A –Supplementary data Video. 1a and 1b).

We then resolved the 3D bud neck structure using time-lapse AiryScan microscopy (Huff, 2015) to image the fluorescent reporter sfGFP-Hof1. The 3D structure showed the characteristic bud neck structure which consists of two parallel doughnut-like septin rings (Fig. 7B- Supplementary data Video. 2). The measured average dimension of bud necks were 0.57 µm in thickness along the mother daughter axis and 0.89 µm in apparent diameter (Fig. 7C) consistent with other measurement (Li et al., 2021) and gave an indication of bud neck dimensions that we used to set spatial parameters in subsequent analysis. Measurement of local crowding from either side of the bud neck was performed using a bespoke semi-automatic analysis workflow which automatically generated regions of set width around a user-specified bud neck. Three regions were defined – one for the bud neck itself and one each for the adjacent regions in the mother and daughter cells. A width of 0.5 µm was specified for the bud neck area as indicated by our length and width quantification. In the mother and daughter cells we analysed only 200 nm region immediately adjacent to the bud neck in each cell (Fig. 8A). We additionally separated cells into three categories, one grouping small buds at the beginning of the division process which we defined as buds with an area below 3 µm$^2$. Large buds, with cell volume comparable to the mother cell and which are close to scission event were defined as buds with area above 7 µm$^2$. All remaining bud sizes were defined as medium size (Fig. 8B). For all three categories the daughter cell maintains a higher ratiometric FRET readout through the cell cycle at the immediate region next to the bud neck while there is a significantly lower FRET at the equivalent region in the mother cell. We measured a mean FRET efficiency of 0.263 ± 0.08 (± SD) for the daughter cell and 0.188 ± 0.05 for the mother cell for the small bud, a 28% difference. For medium bud sizes, the buds have a mean ratiometric FRET of 0.285 ± 0.07 compared to 0.190 ± 0.04 for the mother cells a 68% FRET efficiency jump between the mother cell and the daughter cell. For large buds, we find a mean ratiometric FRET of 0.294 ± 0.07 for the daughter and 0.195 ± 0.05 for the daughter equivalent to a 66% jump. Therefore, using our automatic workflow for analysis, all conditions show significant differences either side of the bud neck, highlighting a polarised crowding trend during replication (Supplemental data, Statistic table 1C). We found this to be true for all cells imaged which indicates that the asymmetry between mother and daughter cells is a stable and maintained condition throughout division (Fig. 8B).

We hypothesised that the cytoplasmic region adjacent to the plasma membrane might exhibit higher molecular crowding as surface proteins with cytoplasmic regions are organized in specific domains. To investigate if there were any local crowding difference close to the plasma membrane we analysed regions expressing the crGE sensor immediately next to the cellular periphery. For each segmented cell we also found the average ratiometric values inside a virtual ring of 200 µm thickness at the periphery of the membrane (outer ring) versus the rest of the cells signal (without the outer ring). This analysis showed non-significant differences between the outer ring, the cell area excluding the outer ring and the whole cell measurement (Fig. 8C – Supplemental data. Statistic table. 1D). The overall region near the membrane therefore appears to experience crowding equivalent to the rest of the cytoplasmic space in the cell. Although the cell membrane is known to be a crowded and highly trafficked region, these dynamics happen on a millisecond timescale and from these data it appears that only highly constrained environment such as the bud neck can sustain a measurable difference in crowding.

**Vacuole inheritance during replication imaged with FM4-64**

To assess if these bud neck crowding events correlate with organelle inheritance, we used time lapse microscopy to follow the inheritance of the vacuole, which is conveniently both a large organelle that is inherited early in the budding process (Li et al., 2021a) . Using a mNeonGreen tagged version of the uracil permease Fur4, which localises to the plasma membrane and also the vacuolar lumen (Paine et al., 2021), we could track inheritance over time (Supplemental Video 3). We also labelled the vacuole by performing a pulse-chase with media containing the red-fluorescent dye FM4-64 (Vida & Emr, 1995). This dye does not diffuse freely through the plasma membrane but instead gets internalised by endocytosis and stains the yeast vacuolar membranes (Fischer-Parton et al., 2000; Petranovic et al., 2010). By tracking the inheritance of FM4-64 labelled vacuoles over time in cells co-expressing either the polarised v-SNARE protein Snc1, or the bud neck marked sfGFP-Hof1 strain we observe vacuole inheritance events. The process observed can be decoupled in three steps. First, the initial vacuole deformation forming an apparent protrusion migrating toward the bud neck region. Then, following by a crossing event, where the vacuole starts occupying the bud neck region and cross the bud neck cell with progressive transport and relaxation toward the daughter cell cytoplasm. Finally, the scission event occurs freeing almost instantly at the bud neck site (Fig. 9 and Supplemental Videos 4a and 4b). We also noticed some rare events of maternal vacuole retraction at the earlier stage of the crossing phase, and thus failure of efficient vacuolar inheritance during these experiments (Supplemental Video 5a and 5b). Stable and timely occupancy of the near bud neck region appears therefore critical to undergo vacuolar migration, this suggests the existence of an adaptive transition before organelles engage crossing. This hypothesis is consistent with our crowding observation at the bud neck. The crowding differential between the two regions boarding the bud neck, in the mother and in the daughter cell, could influence and disturb organelles diffusivity.

**Simultaneous crowding sensing and vacuole visualization during cell division: combined crGE and FM4-64 labelling**

We developed a three-color experiment where the crGE sensor is expressed in cells labelled with FM4-64. We optimized imaging settings so that FRET signal is captured first using the FRET emission filter and excitation condition (see Materials and Methods). Following this we performed an immediate acquisition of FM4-46 exited by a 561 nm argon laser. Fluorescence micrographs and excitation/emission spectrum for these experiments are shown Figures 10 A and B. We tested this set up for cells exposed to strong osmotic stress to confirm we have not impaired crowding sensing quality of our crGE sensor, analysis showed we maintained a statistically significant increase of FRET efficiency between the two conditions, reflecting the typical crowding response occurring upon osmotic stress (Fig. 10C and Supplemental data, Statistic table 1E).

Secondly, we optimized cell growth synchronization of wild-type MATa cells to focus on cell division events during crowding analyses. Cell cycle arrest at the G1 phase of the cell cycle was performed by incubating cells in 10 µM α-factor for two hours (Marina Robles et al., 2017; Udden & Finkelstein, 1978), leading to the forming of so-called shmoos (Merlini et al., 2013) (Fig. 11.). After exchanging buffer to one without α-factor, we observed a return to cell division (Fig. 11B and Supplemental data Video. 7). After eight hours growth time we confirmed that all cells were back to a budding elliptical shape but conserving cell cycle synchronicity across the population (Fig. 11C). We noticed that four hours post α-factor release some cells were still dividing from a shmoo-like phenotype, forming an unwanted subpopulation of cells which were dividing but without forming a regular bud neck. We believe this method is valuable in the field to study yeast dynamics in a synchronized population as it can be applied to any blue-yellow FRET sensor or could be used with green and red fluorescent reporter tags for protein localization studies.

**Tracking single FRET reporters *in vivo***

Finally, we used our Slimfield microscope to track individual crGE2.3 which were endogenously expressed in the same way as the crGE. The two sensors are identical except for the fluorescent proteins used, with crGE2.3 making use of mEGFP and mScarletI as donor and acceptor respectively to make use of the greater single-protein intensities compared to mCerulean3 and mCitrine (Mouton et al., 2020), with the CrGE2.3 thus being suitable for single-molecule tracking. Here we used our redeveloped single molecule tracking code PySTACHIO (Shepherd, Higgins, et al., 2021b) which performs two-channel tracking as well as colocalization analysis using overlap integrals alongside straightforward distance cutoffs as described in the Methods section. The localization was used to calculate the normalized fret NFRET which is possible for this FRET pair due to low spectral overlap and cross-excitation. Detected foci and FRET localizations are shown in Figure 12a.

Figure 12b and c show the histogram and box plot of single molecule FRET values taken from yeast cells in both 0M and 1M NaCl conditions. We see that the distributions of the two conditions' smFRET values is highly similar with means 0.17 and 0.16 respectively, in contrast to our whole-cell measurements which clearly show a high shift in FRET under osmotic stress. This can be explained by the photophysics of the system however. Given that FRET pairs bleach asymmetrically we ensure that we are tracking a functional FRET pair by colocalizing the donor and acceptor channels. However, as FRET increases, more energy is transferred to the acceptor from the donor and the donor intensity decreases in that image channel. In general, we are able to localize foci with intensity above *ca.* 0.7 GFP molecules (Miller et al., 2015). With FRET increasing, this limit will quickly be reached, and the donor intensity will drop, reducing the SNR and making localization of the donor fluorophore impossible. In effect we are therefore only sampling the low FRET pairs, not the full distribution. We believe that this is the cause of our semi-anomalous result, though we also note that in general Slimfield image acquisitions take several minutes and therefore some of the yeast cells may have begun to recover. To counteract this in the future it may be necessary to perform single-molecule imaging only on sensors which are tagged to known (semi-)static structures or organelles so that the exposure time can be increased, and to make use of microfluidics so that cells are imaged only immediately after the stress condition is introduced. Here we used an exposure of 5 ms and found FRET colocalizations for only *ca.* 1% of the localized foci. We could not therefore increase exposure significantly in this highly diffusive regime of cytosolic crGE2.3 sensors. However, an increase to e.g., 40 ms would result in better sampling of the system as well as brighter foci for improved tracking and signal to noise ratios. Similarly, using the even brighter CRONOS sensor (Miyagi et al., 2021) which uses mNeonGreen in place of mEGFP would improve sampling efficiency. We note that this explanation of the NFRET similarity is supported by the higher number of high FRET outliers in the 1M NaCl data in Figure 12c – with greater sampling we may observe a greater difference between conditions.

**Conclusion and discussion**

Understanding intracellular crowding dynamics is highly challenging with crowding playing a role with a range of biomolecular processes and therefore difficult to isolate. Recent progress in optical microscopy and the development of FRET crowding sensor now allows us to access real-time quantification of molecular crowding under stress conditions or during replication.

In our present study, we performed a range of complementary characterization using a FRET based crowding sensor to quantify intracellular crowding between daughter cells and mother cells in budding yeast. We show a stable level of crowding between mother cells and daughter cells, irrespective of cell size. No particular trends were identified either between mother and daughter cells as a function of cell area (Fig. 5). This supports the idea that crowding is a critical homeostasis parameter which is closely linked to cell viability and integrity. Previous studies with the FRET sensor (Mouton et al., 2019) tracking macromolecular crowding through replicative aging also showed crowding stability between mother and daughter cells was maintained. Aging and the budding process are both physiological activities with morphological and cytoplasmic transformation generating two distinct groups with many differences such as organelle size and the accumulation of aging and metabolic by-products. Even with these differences however, the overall physical crowding remains stable.

However, in the face of sudden cellular stress changes in cell morphology or cellular content affects crowding and cell integrity. To test crowding recovery under osmotic stress we salt shocked yeast and imaged their crowding readout through time, which we analysed on whole-cell levels with deep learning segmentation but also in a semi-quantitative subcellular way through heatmap generation. Here we see that crowding readout is not uniformly distributed throughout the cytosol but has a range distribution of values with local hotspots and dynamic evolution through time and along cell division (Fig. 6). Complementary whole cell analysis verified however that the population-level dynamics were largely homogeneous.

We finally targeted the local region at the bud neck and found a difference in ratiometric FRET and hence molecular crowding between the mother and daughter cells at the immediate region bordering the bud neck, with the mother having lower crowding readout than the daughter cell. This interesting result suggests a local crowding polarity between the mother cell and the daughter cell at the bud neck. We hypothesize that this is a consequence of content packaging in the daughter combined with its small volume in expansion. One can however hypothesize a greater role for this observed crowding difference, such as being an effective diffusion barrier during cell division ensuring that only actively transported cargoes reach the growing bud. We also noticed at the bud neck FRET efficiency values are intermediate between the immediate region in mother cells and the one in the daughter cell, which can be interpreted as a local crowding gradient. We therefore speculate crowding in this region to be a stable active marker of the mother/daughter polarity maybe contributing to actively limit the diffusion of freely diffusing material from the mother cell to the daughter cell for example limiting the free diffusion of aging harmful component such as protein aggregates, and other aging by product (Supplemental fig. S2). As opposed to metabolically transported complexes such as the vacuole inherited from the mother cell to the daughter with acute organization and control, mainly via polarised cytoskeleton dependant transport processes involving molecular motors (Fig. 9). Noticeably, essentials cellular compartment are reported inherited via metabolic active transport such as the nucleus (Spichal & Fabre, 2017), the endoplasmic reticulum (Du et al., 2004) and even mitochondria (Boldogh et al., 2001) shown to interact with the cytoskeleton. These observations supporting the presence of a diffusion barrier at the bud neck sorting elements viable to enter in the daughter cells or to retain in the mother cells, where we effectively report locally a stable crowding profile between the two cells.

Globally, the presence of various cellular components and organelles progressively trafficked and accumulating in the daughter cell volume are to consider a possible driven force for the crowding gradient observed. Compartment such as mitochondria, and liposomes shows distinct densities if compared to the cytoplasm (Wang et al., 2014), with distinct shape, size and mobility. If the connection between local crowding dynamic and physical properties within organelles (e.g., density, composition) was not clearly identified and elucidated in the field. Our analysis and reflexions however converge to consider occupancy rate and physical presence close to the bud neck as the main influence on local crowding in the cytoplasm and the diffusion of other molecules and elements in the area.

Organelles are shown inherited in a timely manner, vacuole and mitochondria are inherited at a similar stage of cell division (Li et al., 2021b) with other compartment being inherited earlier such as peroxisomes when the bud start growing or inherited later such as the nucleus at the end of division. The crowding profile we report between the mother cell and daughter cell is also constant throughout cell division, within the three bud size categories measured and representing different state of the cell cycle (Fig.8). This observation supports an effect via element occupancy between the two cells. Various trafficking events being constant and spread through the entire time of cell division. We could also suspect the implication of the recruitment and production of proteins and complexes maintaining the membrane structure of the bud-neck double ring its formation to the typical contractile septin ring split.

As part of our overarching aim to correlate molecular crowding and organelle trafficking events during cell division we have also presented here our latest methodological progress to read out macromolecular crowding while simultaneously visualizing the vacuole during cell division (Fig. 7), demonstrating compatibility for a three-color imaging experiment where crGE (cyan mCerulean3 donor and yellow mCitrine acceptor) conserves its sensing properties when the vacuole was labelled with red dye FM4-64 (Fig. 8). Additionally, we have synchronized yeast cells using α-factor to arrest cell division at the G1 stage, and optimized conditions to ensure that no cells remain in a shmoo state after α-factor release

(Fig. 10). Synchronizing the cell populations will be a crucial step for future experimentation to correlate organelle inheritance events with molecular crowding, and more generally to follow cycle timed events and investigate crowding dependant diffusion barrier between mother cell and daughter cell.

Lateral diffusion barriers in budding yeast have mainly been described as membrane dependant, at the bud neck it was shown that stressed ER confinement in the mother cell was Bud1-GTPase and sphingolipid dependent (Clay et al., 2014). Meanwhile other studies have shown that a diffusion barrier does not exist for the vesicle membrane receptor Snc1 which is locally retained via diffusion at the daughter cell membrane independent from the bud neck structure ((Valdez-Taubas & Pelham, 2003) - Fig. 9). Very little is known about these diffusion barrier dynamics in cells and how they are regulated. It however raises lots of questions and hypotheses regarding crowding modulation at the plasma membrane and cooperation between plasma membrane and cytoplasm dynamics. These powerful and complex processes tied to cellular organization and localization dynamic are closely linked to cellular fate. With new technology and sophisticated fluorescent physicochemical sensors we anticipate that these methods will enable us in the near future to unravel details of the physics of life at the single-molecule level (Leake, 2013), especially when used for emerging correlative microscopy approaches that allow multiple orthogonal data types to be acquired (Leake, 2021)including potential molecular orientational information from super-resolved fluorescence polarization microscopy (Shepherd, Payne-Dwyer, et al., 2021). This may ultimately enable the crucial interplay of trafficking, diffusion, and crowding to be finally elucidated.

**Materials and methods**

Cell strains used:

Yeast cells harbouring a stable integration of the mCerulean3/mCitrine FRET crGE sensor at the *HIS3* locus, expressed from the constitutive *TEF1* promoter were used for molecular crowding measurements as described previously (Shepherd et al., 2020a). The B4741 strain from the genome SWAp-tag yeast library were used to visualise the bud neck, specifically sfGFP-Hof1, sfGFP-Cdc1, sfGFP-Cdc2; sfGFP-Myo1 expressed from the *NOP1*-promoter (Weill et al., 2018).

Cell culturing:

Yeast expressing the FRET sensor were grown in synthetic drop-out media lacking Histidine (2% glucose, 1x yeast nitrogen base; 1x amino acid and base drop-out compositions (SD -His, Formedium Ltd, UK). Cultures were grown to mid-log phase ($OD_{600}$ = 0.4-0.6) prior to harvesting to preparation for optical microscopy.

FM4-63 vacuole labelling:

Prior to imaging live cells were labelled with 0.8 µM FM4-64, incubated in YPD rich media for 1 hour, and then washed twice with drop-out media prior to a 1-hour chase period in synthetic drop-out media, before sample preparation and imaging.

Yeast cell synchronization

Mid log phase BY4741 cells of mating type "a" were arrested in G1 stage (the shmoo phenotype) (Chen & Davis, 2000), after incubation for 120 minutes with 10 µM α-factor pheromone (Zymo-Research Corp, distributed by Cambridge Bioscience LTD, UK). Cells were spun down and washed twice in synthetic drop-out media and left to grow for up to eight hours before imaging to ensure several replications occurred and no cells remained in the shmoo phenotype.

CrGE *in vitro*

CrGE purified as originally previously describe in the literature (Boersma et al., 2015). phenylmethylsulfonyl fluoride (PMSF) in the lysis buffer was replaced by the same concentration of 4-(2-aminoethyl)benzenesulfonyl fluoride hydrochloride (AEBSF), a less toxic alternative for protease inhibition.

*Sample preparation:*

Cells were imaged either in flow channel tunnel slides using 22x22 mm glass coverslip (No. 1.5 BK7 Menzel-Glazer glass coverslips, Germany) coated with 20 µL of 1 mg/mL Concanavalin A (ConA). Slide preparation was perform as previously described (Shepherd et al., 2020). After washing the ConA with 200ul of imaging media, 20 µL of cells were flowed in, the slide was incubated, inverted for 5 minutes in a humidified chamber for adhesion to the ConA coated coverslip and washed again with 200 µl of imaging media, sealed with nail varnish ready for imaging.

Or using 35 mm glass-bottom dishes (Ibidi GmbH, Germany) were first treated with 300ul of 1 mg/mL Concanavalin A (ConA) for 5 min, washed x3 with sterile distilled water and dried in a sterile laminar flow hood. Mid-log phase culture was diluted to $OD_{600}$ = ~0.2 before being adhered to treated coverslips for 5 minutes at ambient temperature and then washed three times with imaging media to remove any unattached cells. Media exchanges were performed using microfluidics as previously described (Laidlaw et al., 2021).

*Confocal microscopy*

Budding yeasts for mother/daughter cells analysis were acquired using a commercial laser scanning confocal microscope (LMS 710 & Axio Imager2, Zeiss) with 1.4 NA Nikon Plan-Apochromat 63x oil-immersion objective lens. mCerulean3 was excited using a 458 nm argon laser at 2.1% of maximum laser power with mCerulean3 emission filter set to 454-515 nm and FRET emission filter set to 524-601 nm. The pinhole size was set to 0.83 Zeiss standard Airy-Units According as in previous work (Shepherd et al., 2020).

Time lapse and three-color microscopy experiments were performed on a commercial laser scanning microscope (LSM880, Zeiss) equipped with an AiryScan module with a 1.4 NA Nikon Plan-Apochromat 63x oil-immersion objective lens. The pinhole size was set to 4.61 Zeiss standard Airy-Units for optimal cytoplasmic signal (equivalent of 3 µm slice/section). Images were taken using the following excitation lasers and imaging wavelength ranges: mCerulean3 458 nm (argon laser) at 1.5% of maximum power and emission filter set to 463/500 nm, FRET 458 nm and emission filter set to 525/606 nm.

For time lapse experiment cells were imaged at 5-minute intervals for time-lapse experiments, including 10 to 15 minutes prior to exchange in standard SD media and for 90 minutes after introduction of stress media (SD supplemented by 1 M NaCl).

For crGE cells dyed with FM4-64, three-color experiments were performed by imaging sequentially the FRET signal as described above followed by a frame mode scanning acquisition exciting FM4-64 with a 561 nm argon laser at 5% of maximum power and emission filter set to 578/731 nm.

The 3D bud neck structure was resolved by AiryScan confocal microscopy, 24 slices of 0.18 µm spacing were acquired. Images were processed for 3D reconstruction with the commercial Zeiss ZEN software and further visualized with the Volume Viewer ImageJ plugin (Barthel, 2005) to generate micrographs.

Image analysis

Analysis of confocal data was performed with a bespoke Python 3 utility based around the YeastSpotter deep learning segmentation tool (Lu et al., 2019). For each detected cell, mean pixel intensities in the FRET and donor channels were extracted to calculate ratiometric FRET values. All plots were generated using the matplotlib (Hunter, 2007) and seaborn (Waskom, n.d.) Python libraries.

For tracking individual cells through time, all DIC images were registered to the first frame using scikit-image (van der Walt et al., 2014). After the first frame was segmented by YeastSpotter, those labels were used for tracking each cell thereafter. To determine if cells were the same between subsequent frames *n* and *n+1*, we found the centre of mass of a given cell in frame *n* and found that pixel in frame *n+1*. If that pixel was within a segmented region in frame *n+1* we assumed the cells to be the same. Though in general this is an unreliable assumption, the surface immobilization by ConA and image registration meant that in this case we found it to be reliable. If cells were pairs between frames, they were given the same label and at the end of the analysis, each labelled region could then be tracked for FRET ratio through time and plotted individually. Here, any cells, which appeared part-way through the acquisition or were lost part-way through the acquisition to drift or surface detachment were neglected – that is, we only analyse through time those cells which were successfully segmented and tracked throughout the full 105 minutes experiment.

For axial analysis of the bud neck, images were first manually annotated in ImageJ/Fiji to define the bud neck area. The line tool was used to draw bud neck lines between mother cells and daughter cell. To identify bud position, lines are drawn in orientated manner: left to right for buds situated in the top part of the image, right to left if bud below the mother cell in the lower part of the image, top to bottom if bud in the right side and bottom to top if on the left. Specifying start and end points here effectively defines a vector, the vector product of which with the *z* axis (0,0,1) gives the vector pointing axially into the mother cell. Annotated images were analyzed with a bespoke Python 3 script. Here, a standard pre-defined area is applied around the bud neckline. From either side of the bud area, in the mother cells and in the daughter cell for a pre-set distance into each cell ratiometric FRET values were extracted. Here to ensure that no noise pixels were included in analysis, we segmented the cells through a Gaussian blur of 3 pixels width followed by Otsu thresholding. This generates an approximate mask which excludes most of the background, but which does not exclude noise pixels in the cells themselves. To do this we did not analyse any pixel with intensity less than the mean background plus three background noise standard deviations as we have reported previously (Shepherd et al., 2020). Again, plots were generated using the matplotlib and seaborn python libraries.

Single-molecule microscopy and analysis

Samples with surface immobilized log-phase *S. cerevisiae* were constructed as described above, using our crGE2.3-incorporating strain described previously (Shepherd et al., 2020). We used our Slimfield single-molecule microscope (Wollman & Leake, 2015) equipped with 488 nm and 561 nm emitting lasers (Obis LS and LX series, respectively) with power set to be around 20 mW at the sample plane for both lasers. We performed all experiments in epifluorescence mode to quickly photobleach most emitters and enter the single molecule regime. We took 5000 frames at 10 ms exposure, and analyzed frames 1000-2000 which we found to have a relatively high number of identifiable foci in both donor and acceptor channels.

Data was analyzed using PySTACHIO (Shepherd, Higgins, et al., 2021b) with snr_min_threshold=0.5 and struct_disk_radius=9 in Alternating Laser Excitation (ALEX) mode. We relaxed our usual constraints on trajectory length because ALEX mode with 10 ms exposure allows considerable diffusion between successive captures in one channel and thus trajectory linking is compromised in this single-molecule regime. Following PySTACHIO analysis, we began by finding the translation-only registration transformation between channels using brightfield images and the pystackreg library (Gregor Lichtner; Philippe Thévenaz, 2021). We then applied the registration transformation to the trajectories identified by PySTACHIO as well as to the fluorescence stacks. We performed colocalization using PySTACHIO with the distance cut-off set to 2 pixels and the overlap integral set to 0.75 – this is the "true" colocalization metric, with the distance cut-off used to decrease computational cost by not calculating overlap integrals for foci which will fall beneath the overlap integral threshold. If spots in the donor and acceptor channel were accepted as being one FRET sensor we then estimated the FRET position using the mean position of the donor and acceptor, and found the summed intensity with local background correction as in PySTACHIO. Finally, we measured the normalized FRET parameter NFRET:

$$\text{NFRET} = \frac{I_{FRET}}{\sqrt{I_D I_A}}$$

which has been reported as a crowding proxy previously (Mouton et al., 2020). These values were plotted using matplotlib (Hunter, 2007). Data analysis here was performed using a bespoke Python routine which made use of scikit-image (van der Walt et al., 2014), Pillow (Clark, 2015), numpy (Oliphant, 2010), and openCV (Bradski & Kaehler, 2000) for image data handling.

**Funding sources**

This project has received funding from the European Union's Horizon 2020 research and innovation programme under the Marie Skłodowska Curie grant agreement no. 764591 (SynCrop), the Leverhulme Trust (reference RPG-2019-156), and BBSRC (reference BB/R001235/1) and Wellcome Trust and the Royal Society grant no. 204636/Z/16/Z.

**Acknowledgements:**

Arnold J Boersma (DWI-Leibniz Institute for Interactive Materials, Aachen, Germany) and Bert Poolman (European Research Institute for the Biology of Ageing, University of Groningen, The Netherlands) for crGE plasmid donation. Maya Schuldiner (Weizmann Institute of Science, Rehovot, Israel) for the shared SWAp-tag yeast library.  Dr Payne-Dwyer for assistance with slimfield microscopy, and Karen Hogg, Grant Calder, Graeme Park, and Karen Hodgkinson (Bioscience Technology Facility, University of York) for support with confocal microscopy.

**Figure captions**

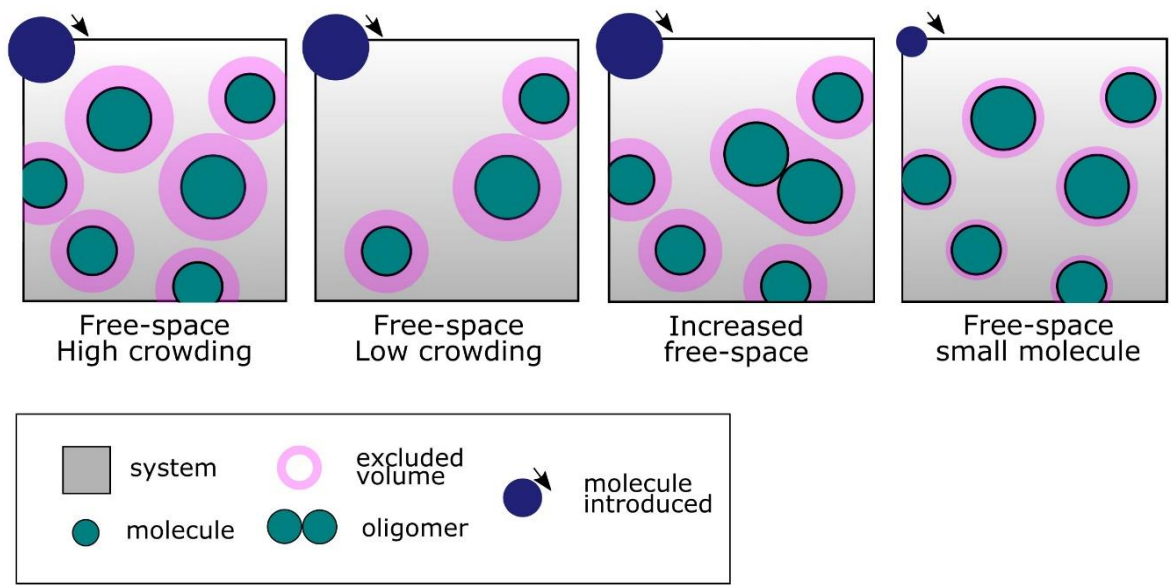

Figure 1: Excluded volume effect, molecules in a confined space.

Illustration of the concept of excluded volume, solvent availability and molecular crowding for molecules in a confined environment. In pink the excluded volume for each molecule in the system therefore limiting mobility and solvent accessibility for the molecules introduced (dark blue molecules on top). The third panel shows how protein oligomerization may modify solvent availability for another molecule and last panel shows the different excluded volume observed in the system for a molecule introduced of a smaller size, excluded volume is dependant of its shape and size.

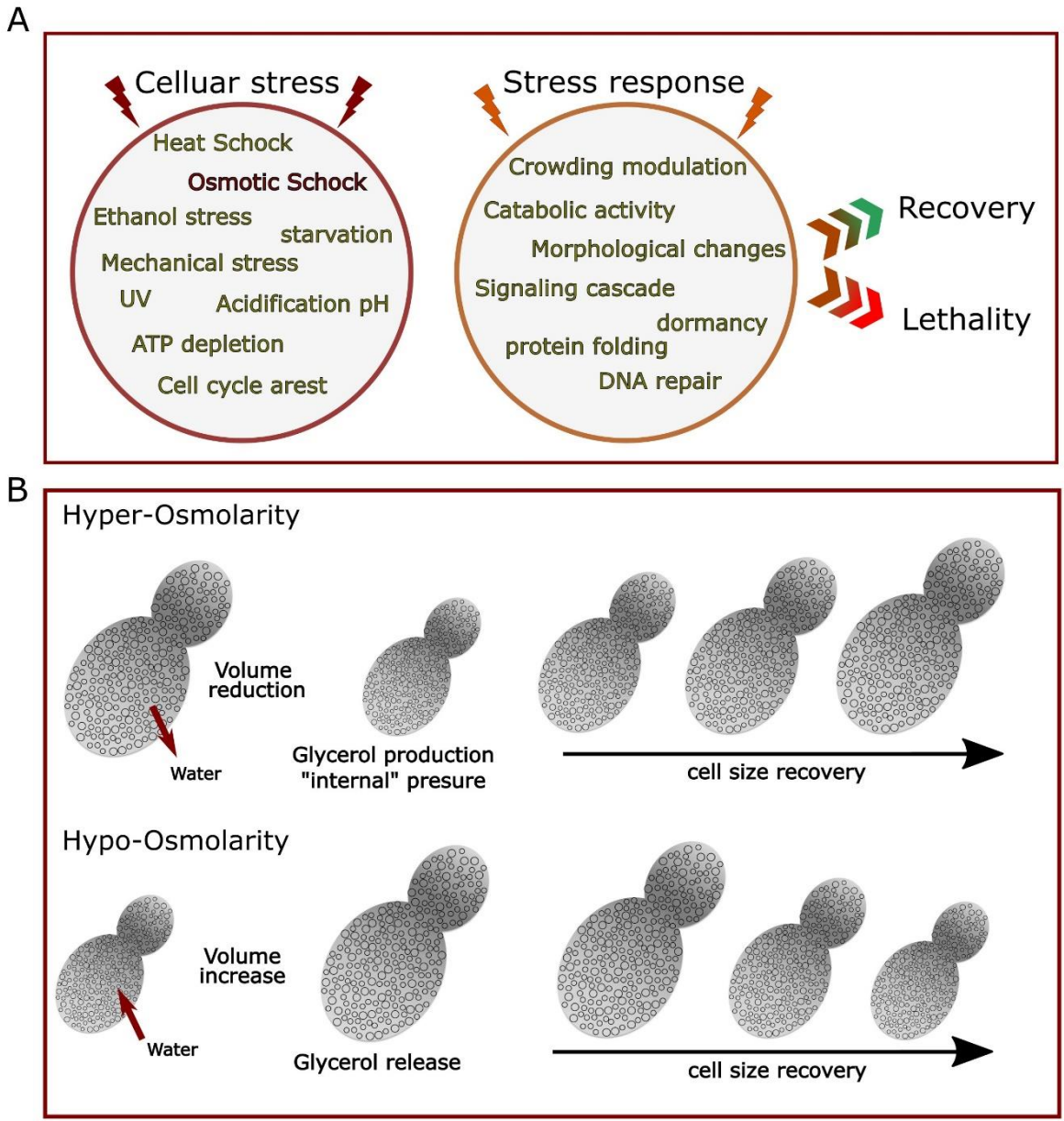

Figure 2: Cellular stress and stress responses

A) General listing of main cellular stress (left) and stress responses (right) that can either trigger a set of metabolic adjustment leading to cell survival and recovery or, if the stress cannot be resolved, leads to cellular death mainly via apoptosis.

B) Schematic osmotic shock response in yeast. Under hyperosmotic stress, cell size drastically reduces as water flows out the cells, resulting in an increased crowding effect in the cytoplasm and the activating osmoregulation Hog1 pathways, the subsequent production of glycerol in the cytoplasm pressure back the cell to recover its initial size. On the contrary during hypo-tonic osmotic stress water flows in and the cell volume increase, less glycerol produced and transporters such as aquaporin are upregulated to allow water to flows in and help the cell recover its original size.

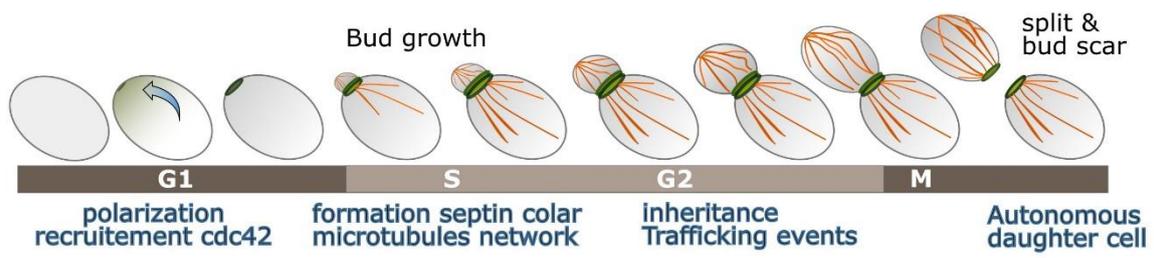

Figure3: Schematic budding yeast polarized cell division

Simple representation of a single yeast cell dividing, polarization to form the bud neck and the septin rings (green) and cytoskeleton polarised through the bud neck in (orange).

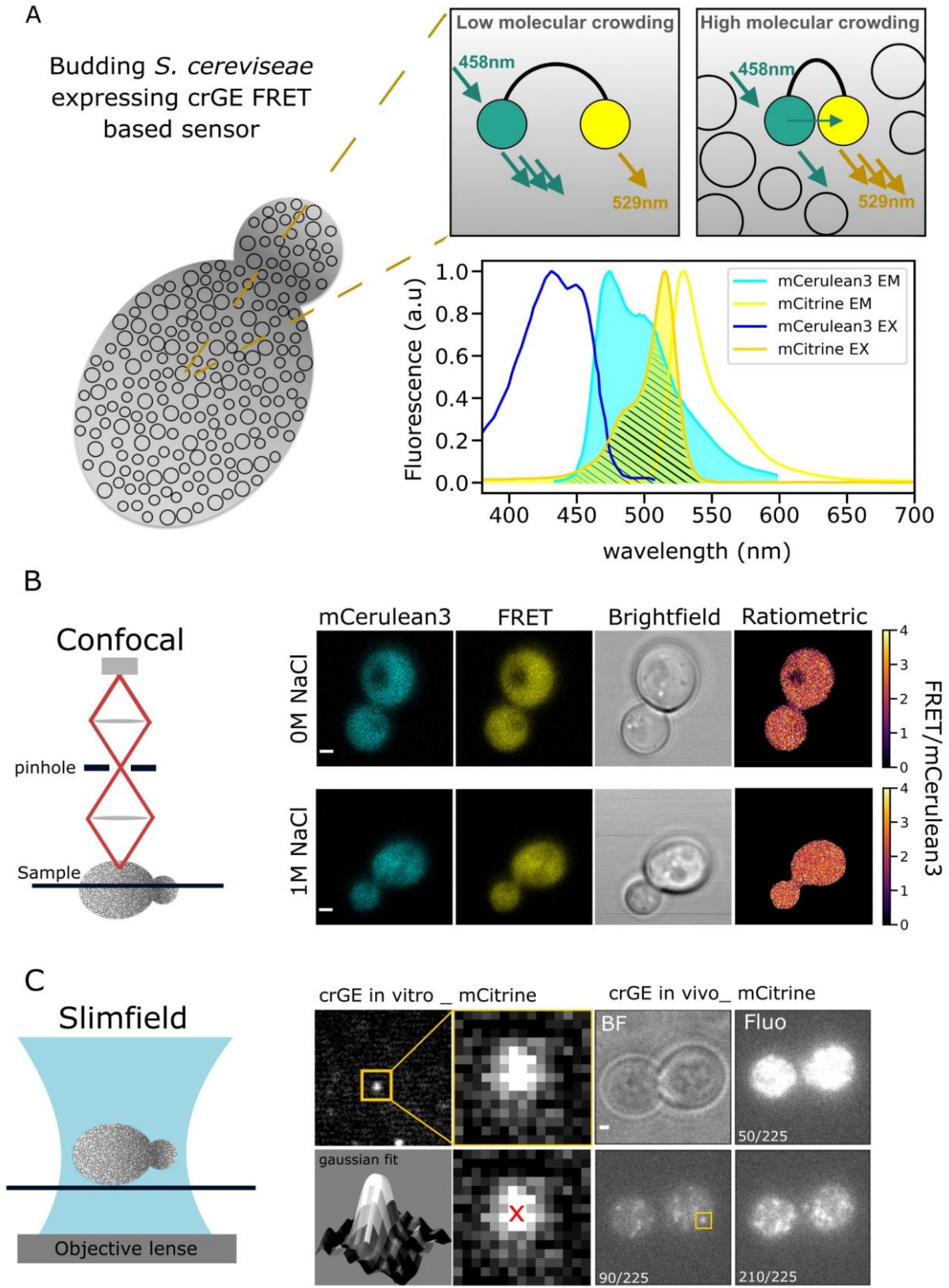

Figure 4: crGE crowding sensor

A) Schematic diagram showing the principles of the crGE FRET sensor. In a crowded environment, the two probes are brought closer to each other which favours the non-radiative energy transfer from mCerulean3 to mCitrine, increasing the FRET efficiency. Inset: the mCerulean3 emission spectrum overlaid with the mCitrine excitation spectrum.

B) Left: schematic representation of confocal microscopy. Right: fluorescence micrographs of yeast cells expressing crGE in both 0 M and 1 M NaCl. Last micrographs on the right show the respective ratiometric maps from the fluorescence images. Scale bar: 1 µm.

C) Left: schematic representation of Slimfield illumination. Right: Visualization of mCitrine single molecules from crGE sensor *in vitro* and *in vivo* expressed in yeast. Iterative Gaussian fits for each spot detected allows us to reach a localization precision around 30 nm.

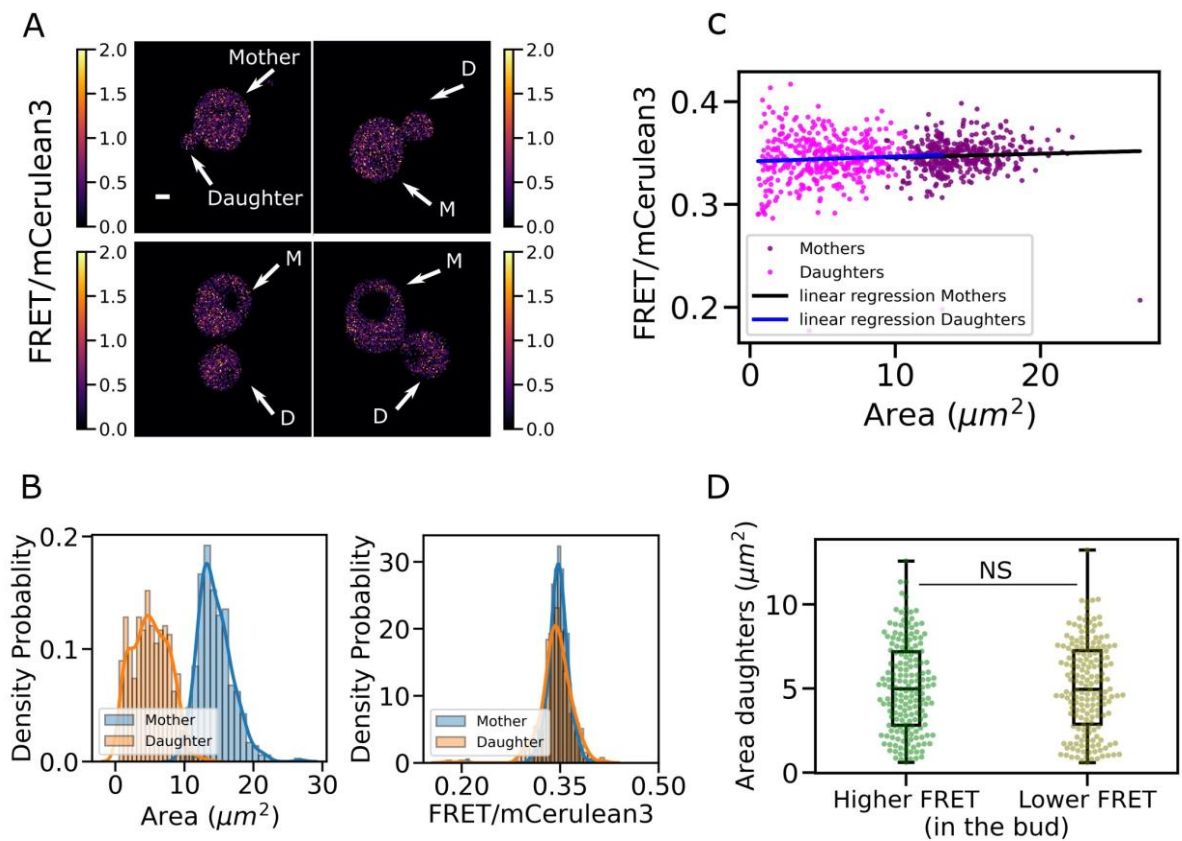

Figure 5: crowding stability between mother and daughter budding yeast cells

A) Yeast cells expressing the crGE sensor. Only budding yeasts were analysed, defined by each mother cell still being attached to its bud or daughter cell as shown in the ratiometric FRET (FRET/mCerulean3) images sample, with white arrows to show the mother cell and the daughter cell of each budding yeast.

B) Scatter plot of mother and daughter cell ratiometric FRET against cell area

C) From left to right, histograms comparing the cell area between mother cells and daughter cells and comparison of the FRET efficiency between mother cells and daughter cells

D) Crowding and cell size dependency in the buds. Two population of cells were measured, one with buds displaying a higher FRET efficiency than their mother cells, the other with buds display a lower FRET efficiency than their mother cells.

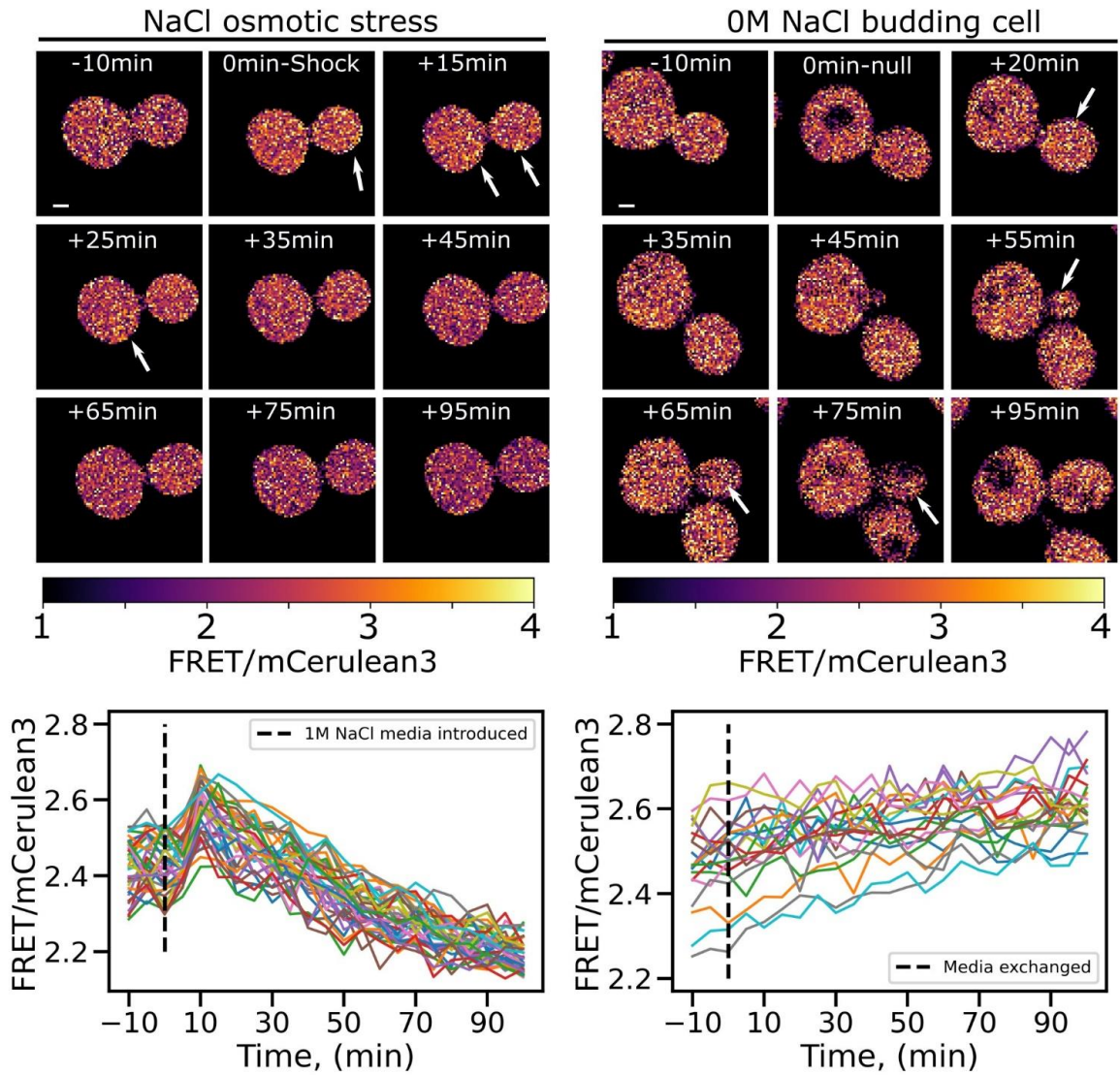

Figure 6: Mapping molecular crowding in single cells

A) Heatmap of the ratio FRET from confocal images acquired during time lapse experiment over 90 minutes. The panel on the left shows cells experiencing osmotic shock with 1M NaCl after 20 minutes and imaged during recovery. On the right panel micrographs show cells left to grow for 90 minutes in non-stress media where a budding event can be observed on the right panel (white arrow)
B) Ratiometric plot through time for each cell revealing the crowding homogeneous behaviour across the cell population.

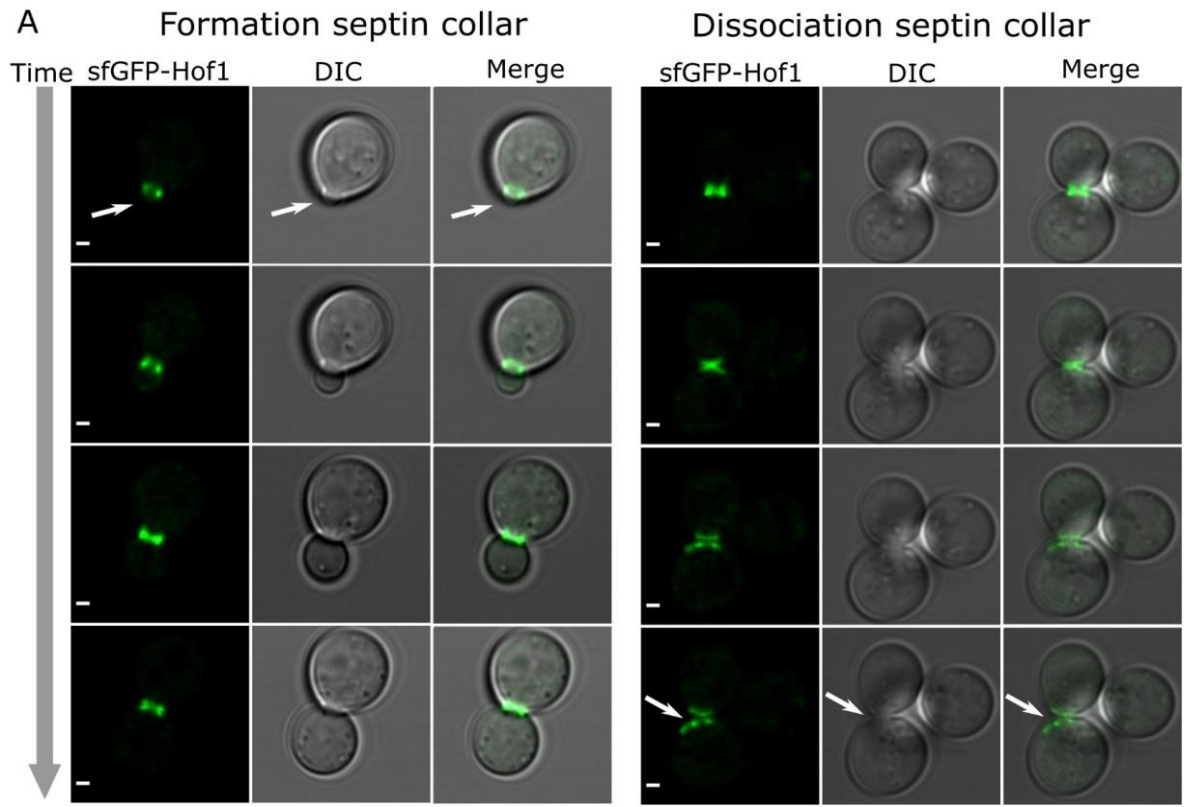

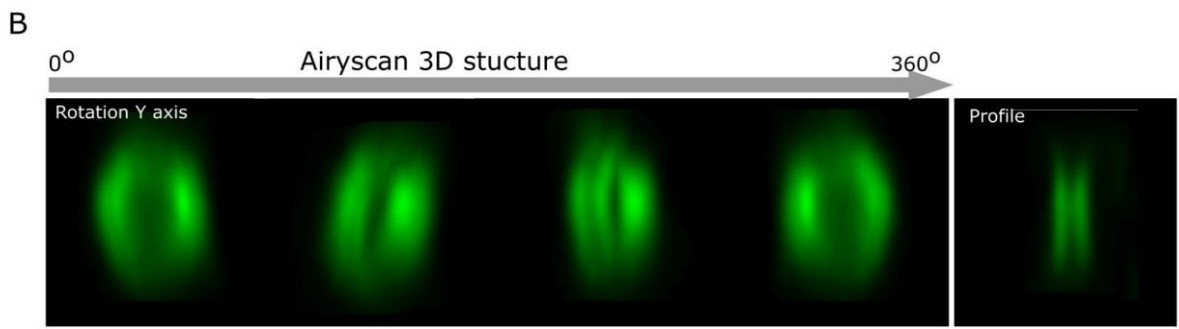

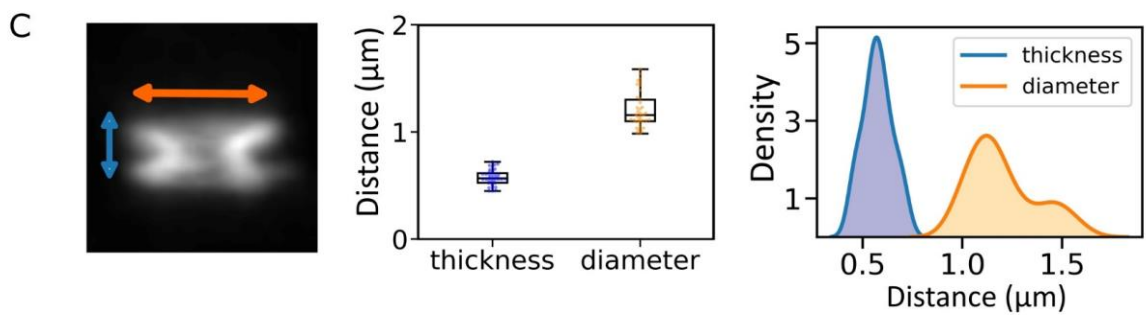

Figure 7: Bud neck imaging

A) Confocal image of sfGFP-Hof1 expressed in budding yeast. Showing the fluorescence channel with sfGFP-Hof1, the DIC grey channel and the merge between the two channels. Scale bar: 1 μm. Right: visualization formation of the bud neck. Left: Visualization cytokinesis event with dissociation of the septin rings- Both indicated with white arrows.

B) 3D structure of the bud neck resolved using Airyscan confocal microscopy, 24 slices of 0.18 µm spacing allowing to fully capturing the bud neck volume. The micrograph shows the 3D volume at different angle of rotation along the y-axis, revealing the doughnut like structure and the profile picture showing the width of the two visible septin contractile rings.

C) Bud neck dimensions. Showing visual of z-projected image of 3D bud neck airy scan stack image (left) with Jitter (center) and KDE plot distribution (right) of the bud neck measured thickness and diameter.

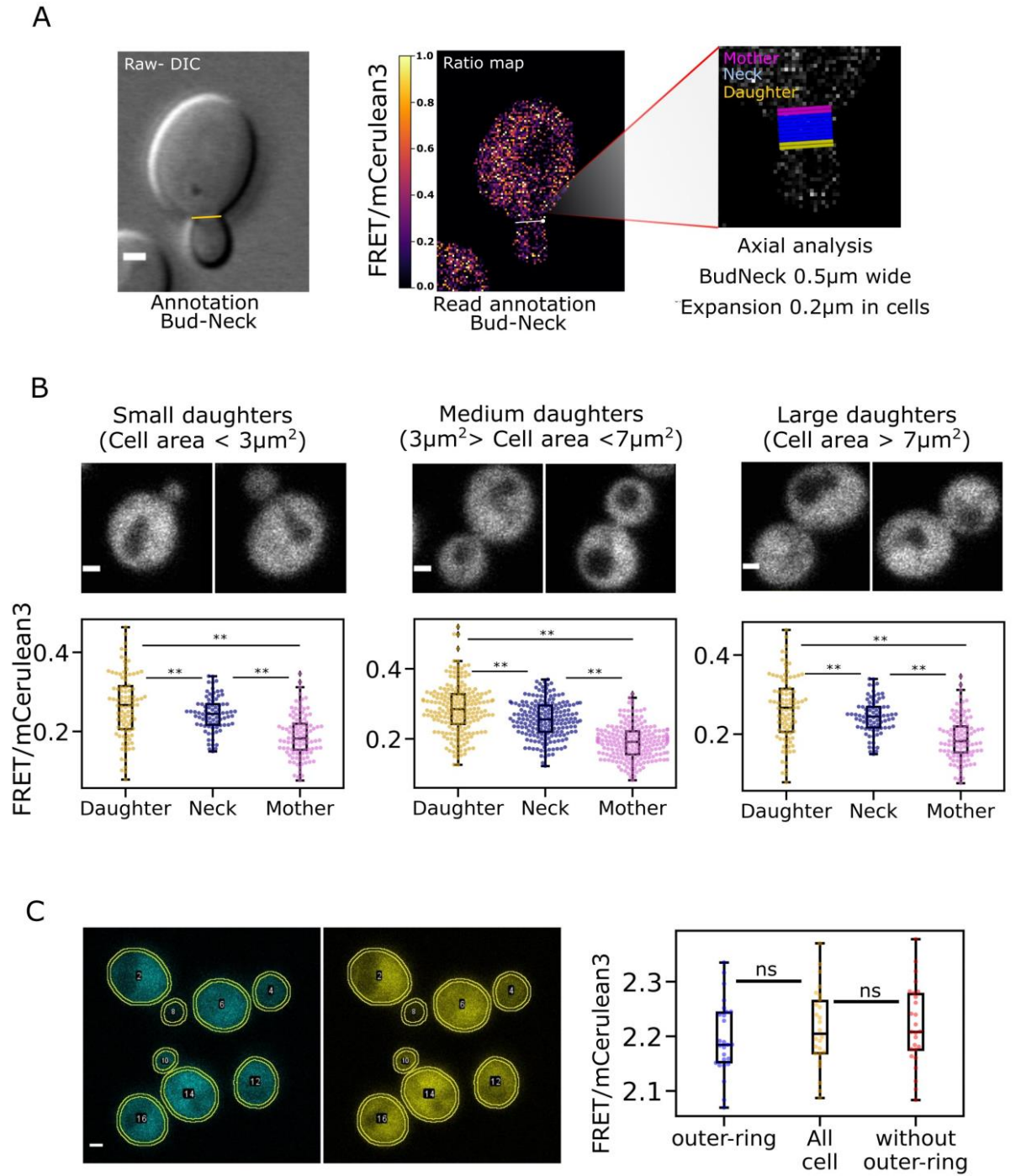

Figure 8: Crowding quantification at localized regions

A) Bud neck annotation, from left to right showing raw image (DIC channel) with manually drawn line between mother cell and daughter cell to define the region of interest (yellow line), Scale bar: 1 μm. The raw annotated image is read by our bespoke python utility, visual in the center showing ratiometric map with bud neck line in white, white dot as the indicator of the line orientation, to determine mother and daughter cell position. In the left output visual of the area measured using our python-based analysis code, in pink and yellow the area of 200 nm, respectively entering the mother cell and the daughter cell, from the defined region of the bud neck in blue.

B) Crowding readout at the bud neck. The figure show an example of each cell category, the small bud with bud size inferior at 3µm$^2$, large buds with area above 7 µm$^2$ and medium category with all the intermediated daughter cell measured. Bellow each category, the respective Jitter plots representative of the FRET efficiency ensured at the bud neck extremity of the mother cells, the defined bud region and the daughter cell. Double asterisk indicates non-parametric Brunner-Munzel test less than 0.005. Scale bar: 1 µm.

C) Average ratiometric FRET in the area underneath the plasma membrane. 200 nm ring at the cytoplasmic periphery of the cell drawn and measured by ImageJ/Fiji macro as presented in the images with segmentation outlines in yellow. Intensity values measured in both channels to calculate FRET/mCerulean3 ratio plotted on the right against the outer ring, the whole cell and the cell without this outer ring region. NS indicate a non-significant correlation between the data with a non-parametric Brunner-Munzel test p value greater than 0.05.

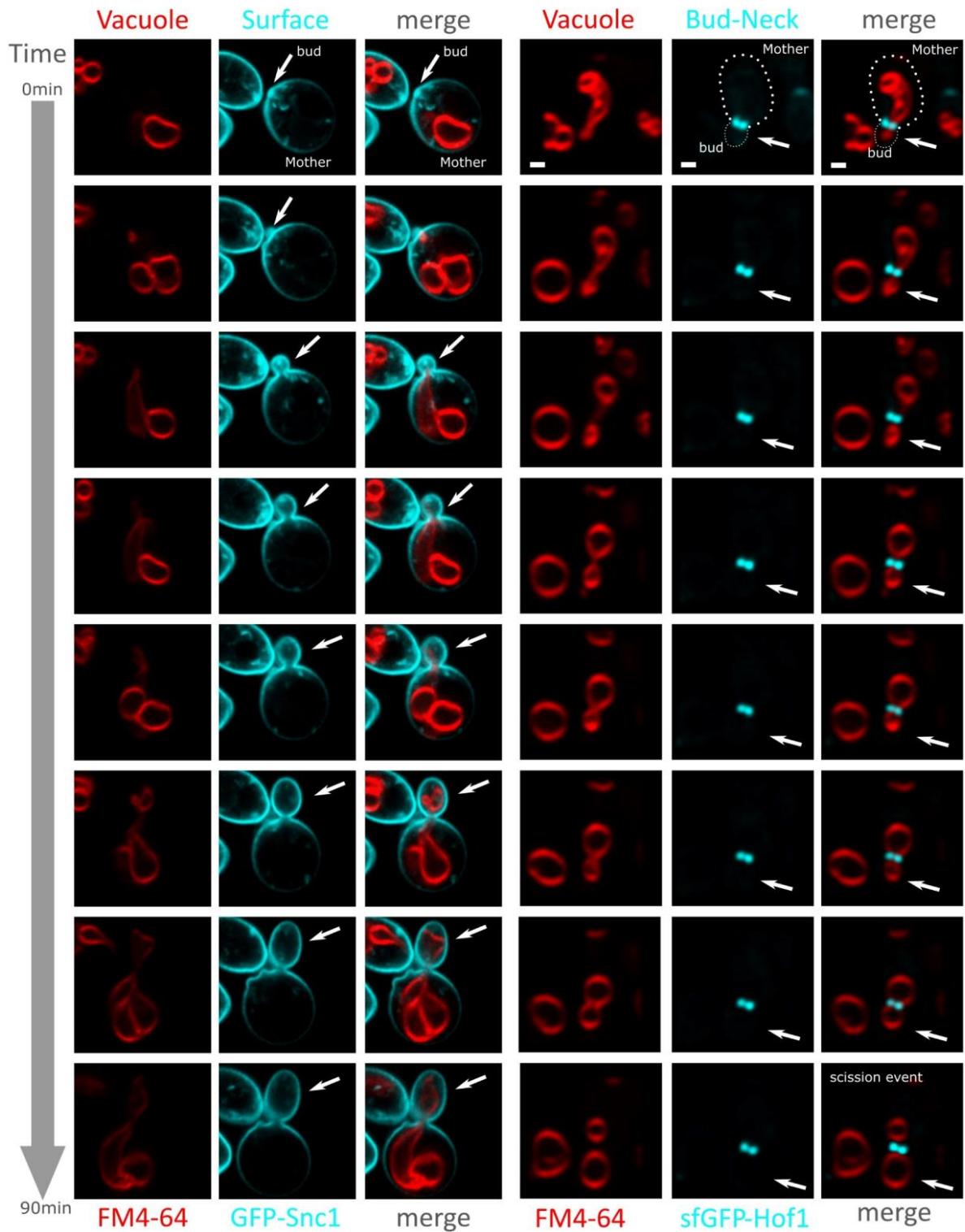

Figure 9: Vacuole inheritance

Strain expressing either GFP-Snc1 bud polarized plasma membrane protein and sfGFP-Hof1 bud neck specific protein, combined with pulse and chase staining with FM4-64 to image the vacuole. Micrographs show each channel for the vacuole in red (FM4-64) and local marker in cyan (GFP & sfGFP), as well as the merge. White arrows to indicate bud/daughter cell position. Scale bar: 1 µm.

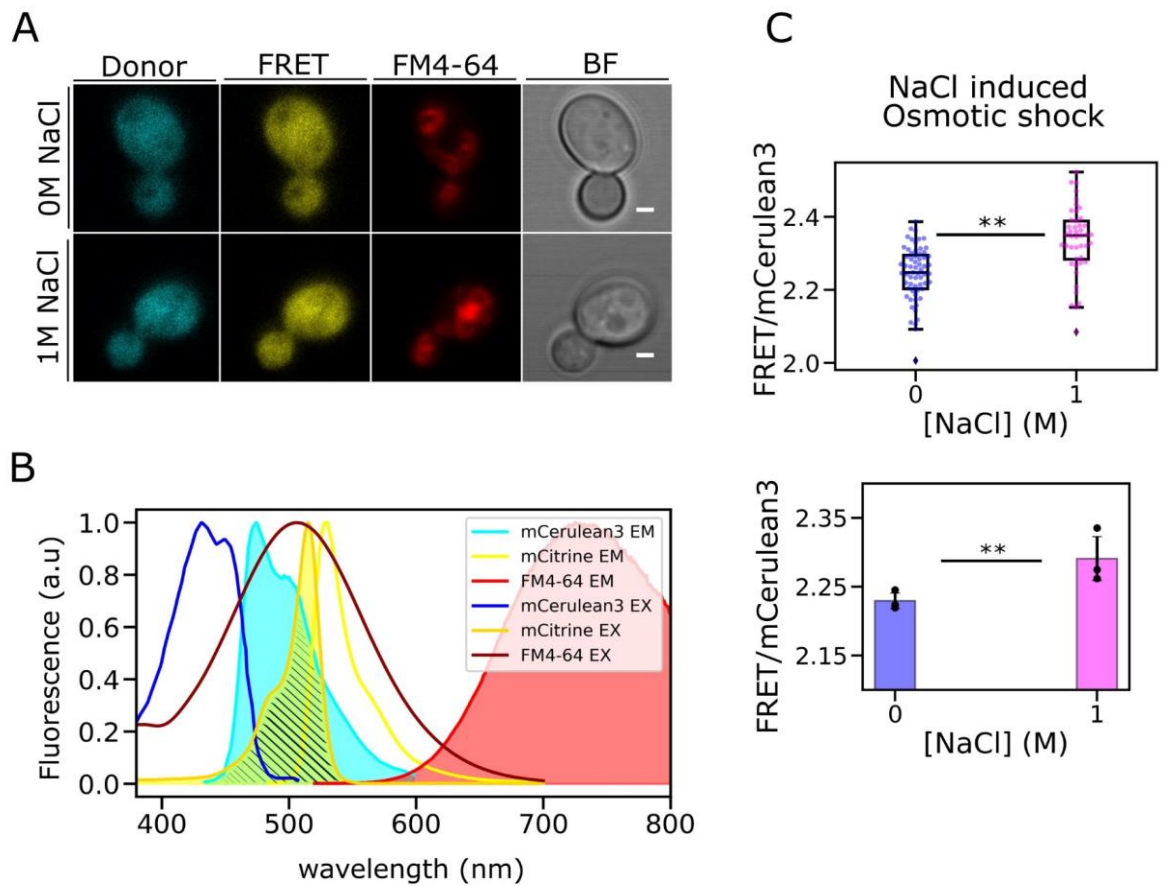

Figure 10: Crowding quantification with simultaneous vacuole imaging

A) Micrographs of three-color imaging for 0M NaCl and 1M NaCl. Scale bar: 1 μm
B) Excitation and emission spectra for mCerulean3, mCitrine and FM4-64M. The excitation laser for FM4-64 is 561 nm which is above the excitation spectra of mCerulean3 (dark blue) and mCitrine (golden yellow). However, acquisition is set so that the FRET signal is acquired before the vacuole marker to minimize impact on the other fluorophore.
C) Quantified crowding for *S. cerevisiae* grown in 2% glucose expressing crGE and labelled with FM4-64, imaged with 0 or 1M NaCl. Inset jitter plot shows a representative crowding dataset for these conditions. Below: box plot of three biological replicates with standard deviation error bar. Double asterisk represents non-parametric Brunner-Munzel test with $p<0.05$.

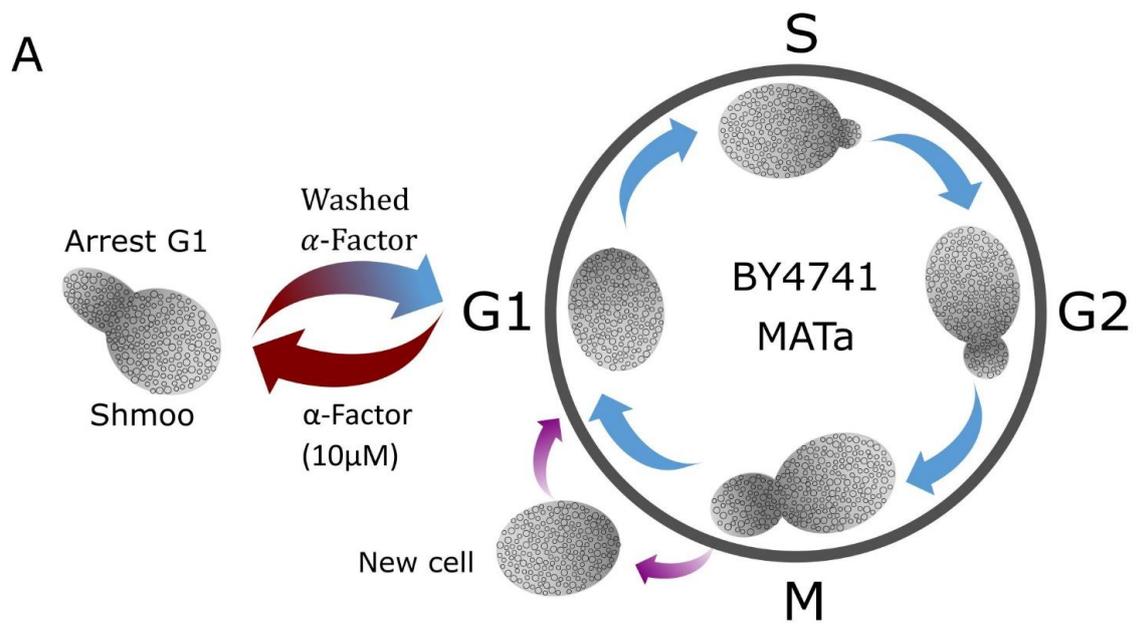
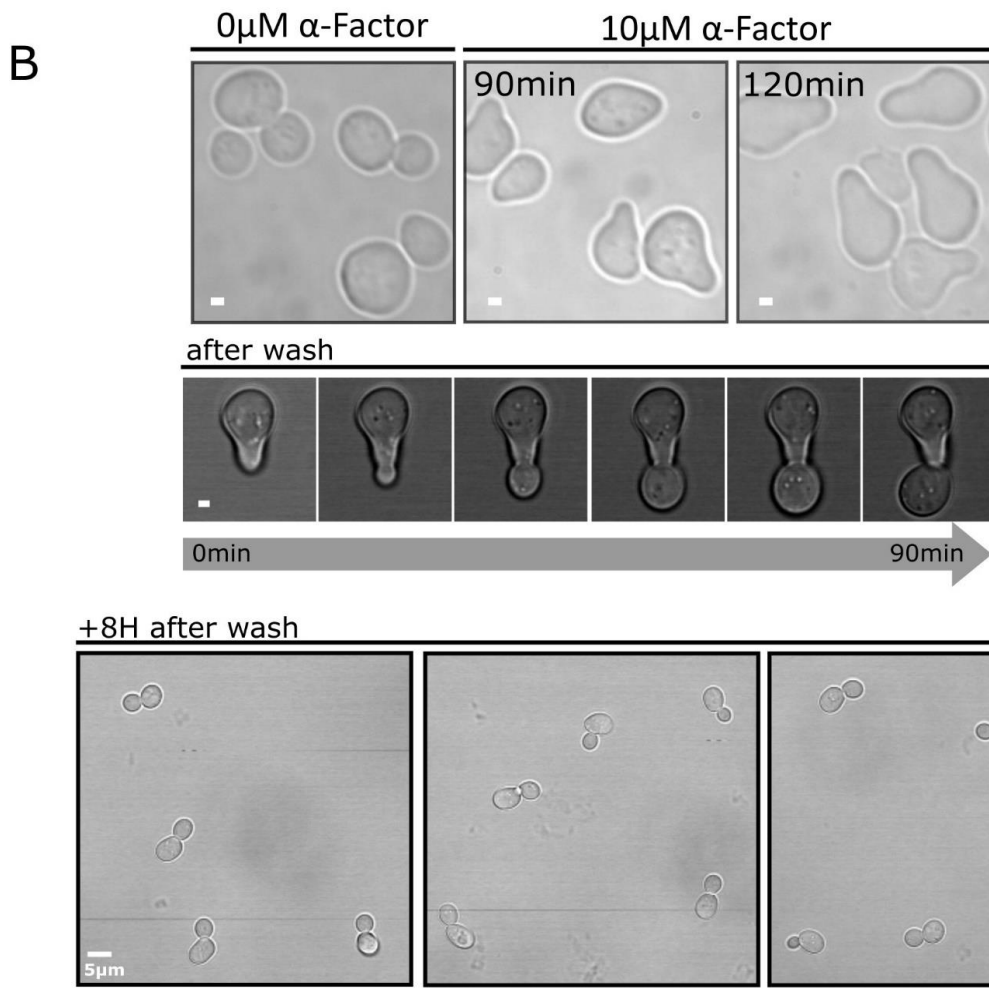

Figure 11: Cell cycle synchronization

a) Schematic representing the cell cycle with its 4 phases (S,G1,M,G2) highlighting arrest of the cycle at G1 phase when MATa yeast cells are exposed to 10 µM α-factor, cells stop dividing and adopted the shmoo phenotype.
b) Brightfield images of shmoos formed after 90 minutes and 120 minutes incubations with 10 µM α-facto (scale bar: 1µm). Micrograph below shows a shmoo entering cell division immediately after removal of α-factor from the media (scale bar: 1 µm), and a micrograph 8 hours after removal showing synchronized budding of yeast cells (scale bar: 5µm).

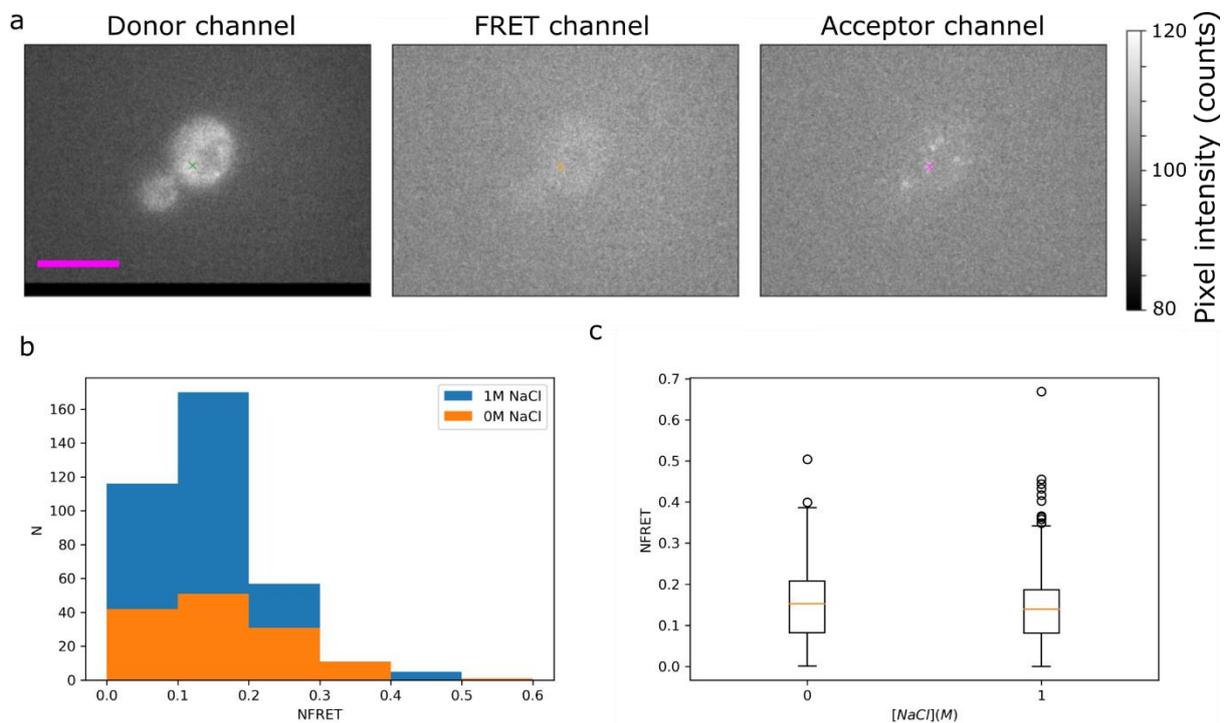

Figure 12: Slimfield microscopy for single molecule detection inVivo of crGE2.3

A) Representative colocalized data from our Slimfield experiments in the three channels post-registration. In the donor channel the localized focus is shown with a green gross, an orange cross is used in the FRET channel to show the average position of the donor and acceptor which is used to measure FRET intensity, and in the acceptor channel we represent the magenta cross shows the localized position of the acceptor. B) Histogram of the NFRET data for *S. cerevisiae* in 0 and 1M NaCl. C) Boxplot of the same data. Bar: 5 µm.

# Supplementary Information

**Investigating molecular crowding during cell division in budding yeast with FRET**

Sarah Lecinski[1,a], Jack W Shepherd[1,a,b], Lewis Frame[c], Imogen Hayton[b], Chris MacDonald[b], Mark C Leake[a,b]*

[1]These authors contributed equally

[a] Department of Physics, University of York, York, YO10 5DD

[b] Department of Biology, University of York, York, YO10 5DD

[c] School of Natural Sciences, University of York, York, YO10 5DD

* To whom correspondence should be addressed. Email mark.leake@york.ac.uk

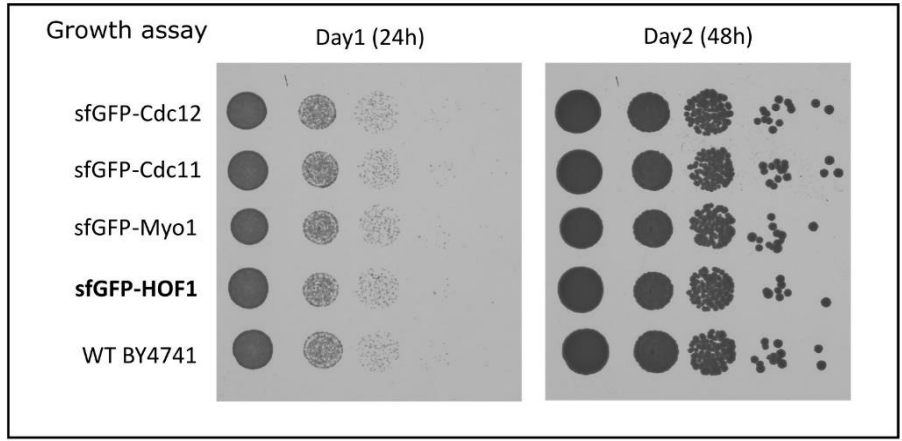

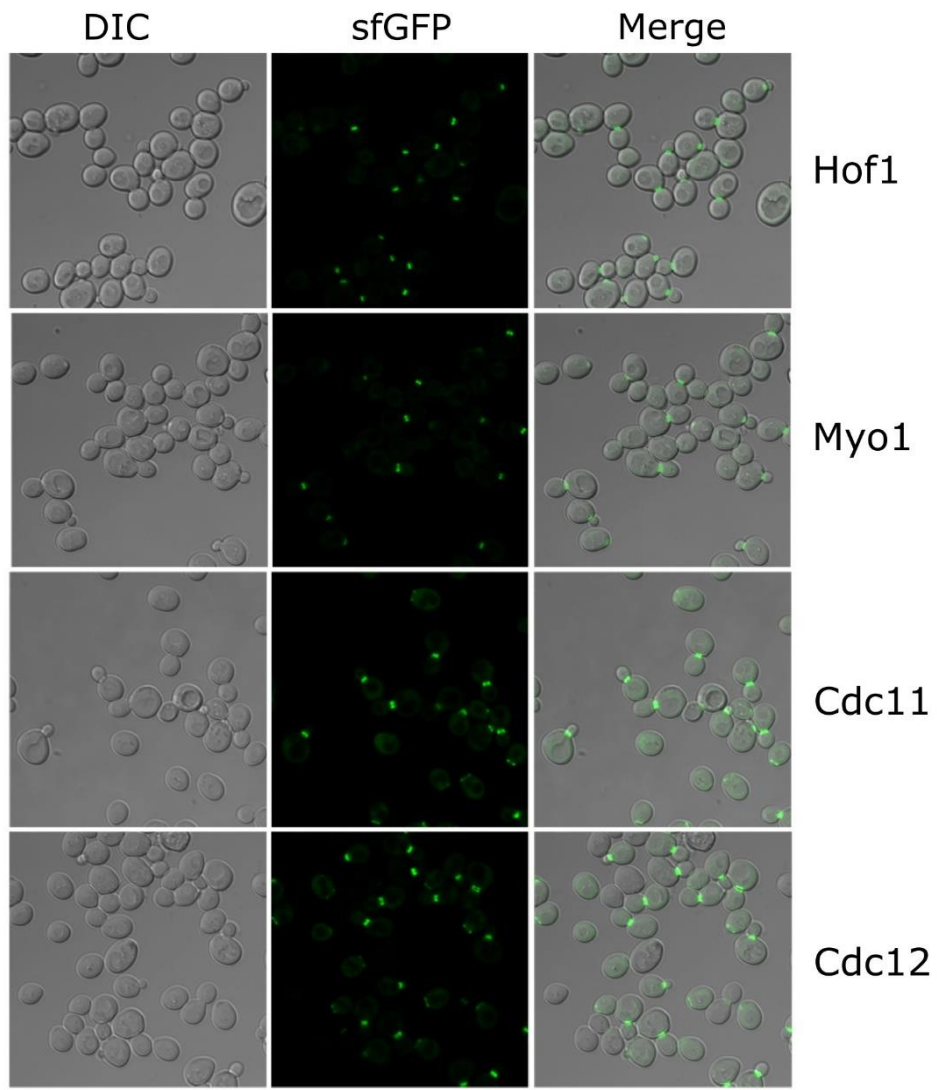

**Supplementary Figure S1: GFP-tagged mid-body markers do not affect growth**

A) Growth Assay for wild type BY4741 (WT) yeast and strains constitutively expressing indicated GFP tagged proteins under control of the *NOP1* promoter in BY4741 background strain. Cultures were grown to mid-log phase, serially diluted 10-fold, plated on minimal media and growth recorded after 24 h and 48 h incubations at 30°C. Strains expressing a fluorescent markers (Hof1, Myo1, Cdc11 and Cdc12) grew similar to the WT strain with no growth defect to report for all strains tested.

B) Confocal imaging was used to localise each GFP-tagged strain, with DIC, GFP and merge micrographs included.

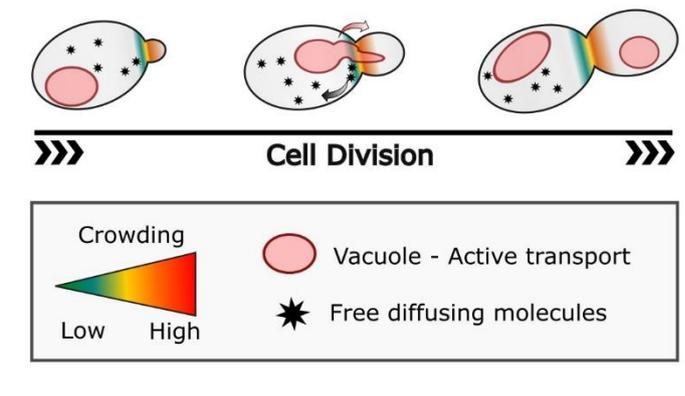

Supplementary Figure S2: Lateral diffusion barrier model for bud neck crowding.

Schematic diagram representing low to high crowding gradient from mother to daughter cells within the 200 nm region bordering the bud neck (Fig. 8). Hypothetic model for lateral diffusion barrier between the two connected cells depicted with colours, promoting the retention in the mother cell of free diffusion particles (black stars) while essential organelles such as the vacuole (in pink) can migrate to the daughter cell. In grey the rest of the cell where crowding profile of distribution not represented.

Supplemental statistic tables

The following tables shows all statistical tests performed for dataset presented and discussed in this study. All p values calculated using the non-parametric Brunner-Munzel test.

A) Comparison between the mother and daughter cell population, area and ratiometric FRET

| Comparison mother cells daughter cells | Mother | Daughter |
|---|---|---|
| Mean area | 14.453 | 5.124 |
| Mean ratiometric FRET | 0.347 | 0.344 |
| Ratiometric FRET standard deviation | 0.015 | 0.023 |
| Mother-daughter area comparison p value | 0 | |
| Mother-daughter ratiometric FRET p value | 0.03 | |

B) Daughter cell crowding and area dependency in budding yeast

| Daughter cell crowding and area dependency analysis | Daughter with higher FRET than mother | Daughter with lower FRET than mother |
|---|---|---|
| Mean area | 5.08 | 5.1 |
| Median area | 4.99 | 4.94 |
| p value | 0.87 ||

C) Local crowding at the bud neck - mother/daughter cell 200 nm region adjacent to the bud neck – three budding category stages (small: bud size smaller than 3µm$^2$ ; medium: bud size between 3µm$^2$ and 7µm$^2$ ; Large: bud size larger than 7µm$^2$ )

| Axial analysis | Small | Medium | Large |
|---|---|---|---|
| Daughter-bud comparison p value | 0.05 | 2.46E-06 | 0.0002 |
| Mother-bud comparison p value | 6.82E-16 | 1.64E-42 | 7.66E-14 |
| Daughter-mother comparison p value | 7.8µE-14 | 1.61E-62 | 7.04E-30 |
| Mean daughter ratiometric FRET | 0.263 | 0.285 | 0.293 |
| Mean bud neck ratiometric FRET | 0.244 | 0.255 | 0.252 |
| Mean mother ratiometric FRET | 0.180 | 0.190 | 0.190 |
| Daughter ratiometric FRET standard deviation | 0.08 | 0.07 | 0.07 |
| Bud neck FRET standard deviation | 0.04 | 0.05 | 0.05 |
| Mother cell ratiometric FRET standard deviation | 0.05 | 0.04 | 0.05 |

D) Local crowding at the cell periphery

| At the cell periphery | Outer-ring | All cell | Without outer-ring |
|---|---|---|---|
| Outer ring/all cell comparison p value | 0.31 | | |
| All cell/cell without outer ring comparison p value | | 0.77 ||
| Outer ring/cell without outer ring p value | 0.24 |||
| Mean Ratiometric FRET | 2.197 | 2.215 | 2.22 |
| Median Ratiometric FRET | 2.184 | 2.204 | 2.208 |

E) Osmotic shock crowding in cells expressing the crGE sensor and labelled with FM4-64

| Combined CrGE & FM4-64 dye 0 to 1M NaCl shock | Replicate 1 | Replicate 2 | Replicate 3 |
|---|---|---|---|
| 0 M/1 M comparison p value | 5.60E-08 | 4.22E-06 | 0.0014 |
| Mean ratiometric FRET shift (%) | 4% | 2.50% | 1.50% |
| Median ratiometric FRET shift (%) | 4.50% | 3.10% | 1.70% |

Supplemental data – Movies

Video 1a and 1b: Microscopy of GFP-Hof1 cells budding (1a) and undergoing cytokinesis (1b)

Video 2: AiryScan confocal microscopy of cells expressing GFP-Hof 1 to define 3D bud-neck structure

Video3: Time-lapse microscopy of wild type cells expressing Fur4-mNG

Video 4a and 4b:  Time-lapse AiryScan microscopy of cells expressing GFP-Snc1 and FM4-64 labelled vacuoles

Video 5a and 5b: Time-lapse microscopy of cells expressing GFP-Hof1 and inherited vacuoles stained with FM4-64

Video 6a and 6b: Time-lapse microscopy of aborted FM4-64 stained vacuolar inheritance event, in cells also expressing GFP-Hof1

Video 7:  crGE sensor expressed in yeast cells arrested and then released by exposure to α-factor